\documentclass[aps,prb,twocolumn,a4paper,floatfix,showpacs,superscriptaddress]{revtex4}

\usepackage{epsfig}
\usepackage{graphicx}
\usepackage{hyperref}
\usepackage{color}
\usepackage{amsmath}
\usepackage[normalem]{ulem}
\usepackage{lineno}

\usepackage{multirow}
\usepackage{upgreek} 

\begin{document}

\title{A phase-field model for simulating hydrogen-induced pitting corrosion with solid-solid phase transformation in the metal}

\author{Jie Sheng}
\affiliation{Laboratory of Computational Physics, Institute of Applied Physics and Computational Mathematics, Beijing 100088, China}

\author{Yue-Chao Wang}
\affiliation{Laboratory of Computational Physics, Institute of Applied Physics and Computational Mathematics, Beijing 100088, China}

\author{Yu Liu}
\email{liu\_yu@iapcm.ac.cn}
\affiliation{Laboratory of Computational Physics, Institute of Applied Physics and Computational Mathematics, Beijing 100088, China}

\author{Shuai Wu}
\affiliation{Laboratory of Computational Physics, Institute of Applied Physics and Computational Mathematics, Beijing 100088, China}

\author{Ke Xu}
\affiliation{Laboratory of Computational Physics, Institute of Applied Physics and Computational Mathematics, Beijing 100088, China}

\author{Zi-Hang Chen}
\affiliation{Laboratory of Computational Physics, Institute of Applied Physics and Computational Mathematics, Beijing 100088, China}

\author{Bo Sun}
\affiliation{Laboratory of Computational Physics, Institute of Applied Physics and Computational Mathematics, Beijing 100088, China}

\author{Hai-Feng Liu}
\affiliation{Laboratory of Computational Physics, Institute of Applied Physics and Computational Mathematics, Beijing 100088, China}

\author{Hai-Feng Song}
\email{song\_haifeng@iapcm.ac.cn}
\affiliation{Laboratory of Computational Physics, Institute of Applied Physics and Computational Mathematics, Beijing 100088, China}

\date{\today}

\begin{abstract}
Hydrogen-induced pitting corrosion of metallic is a common phenomenon that damages the integrity and durability of the materials. Its numerical simulation is still a challenge due to many complex mechanisms, especially solid-solid phase transformation and mechanical interaction, leading to the anisotropic growth of hydride and inducing some bulges on the metal surface. In our work, we propose a phase-field model and numerical technique for simulation of hydrogen-induced pitting corrosion, and apply it to the system of $\alpha$-Uranium. In our model, the elastic strain energy is introduced to approximate the anisotropic pit morphology induced by the mechanical interaction between metal and hydride. For the numerical technique, the free boundary condition based on the finite element method is adopted to introduce the bulges of the metal surface. By the application of our model and numerical technique, the anisotropic pit morphology with a bulge on the metal surface in agreement with experiments of $\alpha$-Uranium is obtained. Moreover, the compression of $\alpha$-Uranium and the dilation of its hydride are discovered, which develops the deep understanding of hydrogen-induced pitting corrosion. This model is expected to be applied to the health detection of hydrogen-induced pitting corrosion of metal in the industry.

\end{abstract}

\maketitle

\section{Introduction}
Hydrogen-induced pitting corrosion, as a common local corrosion\cite{ernst2002pit1,ernst2002pit2}, reduces the integrity and durability of metals and even leads to metal failure\cite{banos2018review}. The hydrogen-induced pitting corrosion is induced by the transformation from metals to hydrides, such as the pitting in uranium\cite{banos2018review} and zirconium\cite{alvarez2011phase}. Hydrogen-induced pitting corrosion usually involves four periods\cite{brierley2016anisotropic}, named incubation period, early growth, oxide cracking, oxide spalling and continued growth. Early growth is a significant period to investigate the morphology and kinetics of hydride and the mechanism of oxide film rupture. In this period, the hydride usually exhibits anisotropic growth\cite{bingert2004microtextural,jones2013surface}. Moreover, the growth of hydride exerts an outward strain and leads to a bulge on the metal surface\cite{ji2019mechanism,bingert2004microtextural,jones2013surface}. The formation mechanisms behind these important features still are unclear, and have been the focus of many researchers\cite{banos2018review}. Some experimental efforts have been put on this topic, for example Ji \textit{et al.}\cite{ji2019mechanism} have tried to ascertain the formation mechanisms of uranium hydride, and Bingert \textit{et al.}\cite{bingert2004microtextural} have carried out a investigation about the effect of the heterogeneity of the microstructure on the growth of hydride in uranium. However, for providing a more detailed physical interpretation, and supporting the design and application of experiments, a numerical simulation is needed in advance. Thus, it requires a computable model and simulation method to accurately describe the anisotropic growth of hydride and bulges of the metal surface.

In the past decades, great efforts have been made in the development of numerical methods for pitting corrosion. According to the treatment of the corrosion interface, these methods could be divided into the sharp interface methods and diffuse interface methods. The former, including the finite volume methods\cite{scheiner2007stable,scheiner2009finite}, the arbitrary Lagrangian-Eulerian methods\cite{sun2014arbitrary} and the level set methods\cite{sethian1996fast}, has been successfully applied to the pitting kinetics simulation of metal-electrolyte corrosion system. However, the needs of tracking interface position according to the movement speed and creating a matching element mesh at each time step\cite{mai2016phase} improves the algorithm complexity and implementation difficulty of these methods. The latter, represented by the phase field\cite{yang2021explicit,yang2020multiphase,mai2016phase,ansari2018phase} (PF), avoids tracking the moving interface by hiding the corrosion interface in the control equation\cite{mai2016phase,yang2021explicit}. The PF method defines the free energy of the system based on thermodynamically consistent and simulates the microstructure evolution of the system by minimizing the free energy of the system\cite{ansari2018phase}. Therefore, it is easy to consider the influence of free energy variation caused by the interaction of internal mechanisms (such as concentration and phase transformation) or external environments (such as stress, strain and temperature) on the structural evolution of the system. Up to now, the PF method has received significant attention for simulating microstructure evolution\cite{li2017review}, such as solidification, dendrite growth, spinodal decomposition, dislocation dynamics, crack propagation, phase transformation and corrosion\cite{mai2016phase,mai2017phase,ansari2018phase,yang2021hydride}. 

Recently, the PF method has been tried to study solid-liquid phase transformation pitting corrosion in the metal-electrolyte system. Mai \textit{et al.}\cite{mai2016phase,mai2017phase} proposed two PF models for simulating the pitting corrosion of stainless steel in the liquid-phase electrolyte. Ansari \textit{et al.}\cite{ansari2018phase} further developed these PF models by introducing electrode reactions. These models also have the potential to simulate the hydrogen-induced pitting corrosion with solid-solid phase transformation. But the lack of elastic strain energy
limits their ability about the description of the anisotropic pit morphology and the bulge on the metal surface, especially in hydrogen-induced pitting corrosion.

Therefore, to better describe the above significant features in hydrogen-induced pitting corrosion, we develop a PF model. In our model, to obtain an anisotropic pit morphology, we couple the elastic strain energy between metals and hydrides into the phase-field model based on Kim–Kim–Suzuki (KKS) assumption\cite{kim1999phase}. The governing equations of model are derived by minimizing the free energy of system during the pitting process. For the bulge on the metal surface, we adopt a numerical technique including the free boundary condition based on the finite element method (FEM).  The validity and accuracy of the model are verified in $\alpha$-Uranium ($\alpha$-U), and the deformation mechanism of material in hydrogen-induced pitting corrosion is proposed.

The remainder of this manuscript is organized as follows. In section \ref{sec2}, we firstly describe the hydrogen-induced pitting corrosion of $\alpha$-U. Then, the proposed PF model is expounded in detail, including the governing equations of the PF model and the total free energy of the system. The total free energy consists of bulk free energy, interface gradient energy and elastic strain energy. In section \ref{sec3}, the morphology, kinetics, kinematics and stress during the hydrogen-induced pitting process of $\alpha$-U are studied by the proposed PF model. Section \ref{sec4} closes the paper with a summary of the main work.

\section{The PF model of the hydrogen-induced pitting corrosion} \label{sec2} 
\subsection{Problem description}
\begin{figure}
	\centering
	\includegraphics[width=0.48\textwidth]{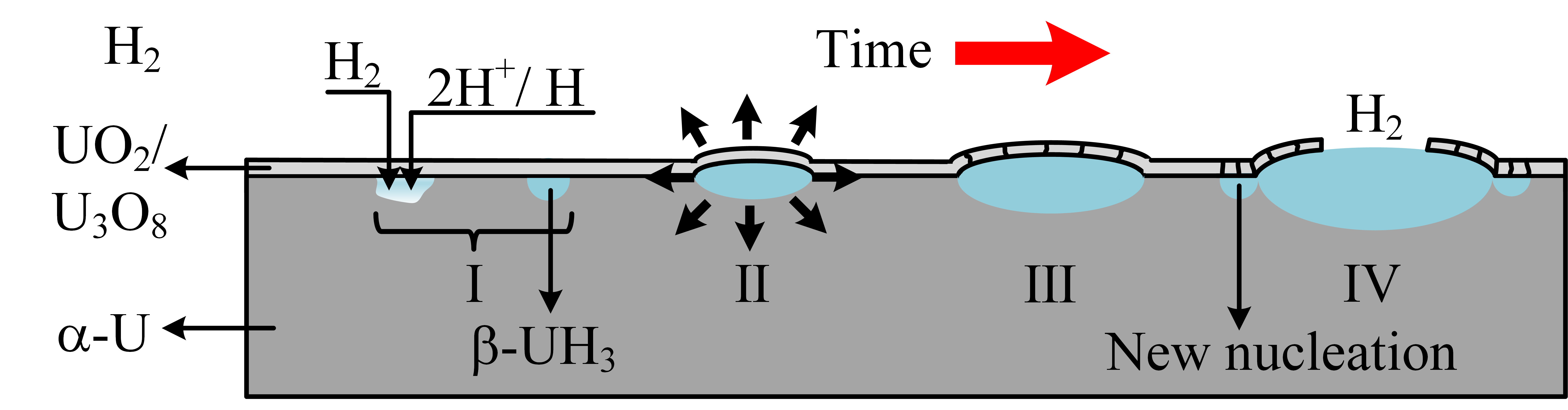}
	\caption{\label{fig1}Schematic diagram of the hydrogen-induced pitting corrosion in $\alpha$-U\cite{banos2018review,brierley2016anisotropic}: \uppercase\expandafter{\romannumeral1}. Incubation period, \uppercase\expandafter{\romannumeral2}. Early growth, 
		\uppercase\expandafter{\romannumeral3}. Oxide cracking, 
		\uppercase\expandafter{\romannumeral4}. Oxide spalling and continued growth.}
\end{figure}

Uranium metal attacked by hydrogen is a typical system of hydrogen-induced pitting corrosion. Fig. \ref{fig1} is a schematic diagram of the hydrogen-induced pitting corrosion in $\alpha$-U at hydrogen atmosphere\cite{banos2018review,brierley2016anisotropic}. The hydrogen-induced pitting corrosion could be divided into the following four periods:

\uppercase\expandafter{\romannumeral1}. Incubation period: Hydrogen accumulates at the oxide-metal interface\cite{banos2018review}, producing an initial hydride precipitate\cite{jones2013surface,scott2007ud3} (usually $\beta$-UH$_{3}$ at the temperature\cite{banos2018review} above 523K).

\uppercase\expandafter{\romannumeral2}. Early growth: Once hydride precipitate has occurred, the early growth period begins. During this period, the initial precipitate grows gradually as the $\alpha$-U reacts with hydrogen at the metal-hydride interface. Meanwhile, the growth of hydride precipitate at the oxide-metal interface exerts an outward strain, related to the dilation associated with hydride formation, and leads to a bulge on the metal surface \cite{brierley2016anisotropic,jones2013surface}.

\uppercase\expandafter{\romannumeral3}. Oxide cracking: Oxide cracking occurs when the outward strain of the hydride is sufficient to distort the oxide film on the surface to rupture. 

\uppercase\expandafter{\romannumeral4}. Oxide spalling and continued growth: When the oxide film is broken to a certain extent, it causes the oxide film to flake off\cite{jones2013surface,banos2017investigation}, and the hydrogen diffuses directly into the hydride, accelerating the growth of existing hydride precipitate and the generation of new hydride precipitates\cite{brierley2016anisotropic}.

In our work, we focus on the growth of an existing hydride precipitate in the \uppercase\expandafter{\romannumeral2} period, and formulated and implemented a PF model to study morphology, kinetics, kinematics and stress of this period.

\begin{figure}[b]
	\centering
	\includegraphics[width=0.48\textwidth]{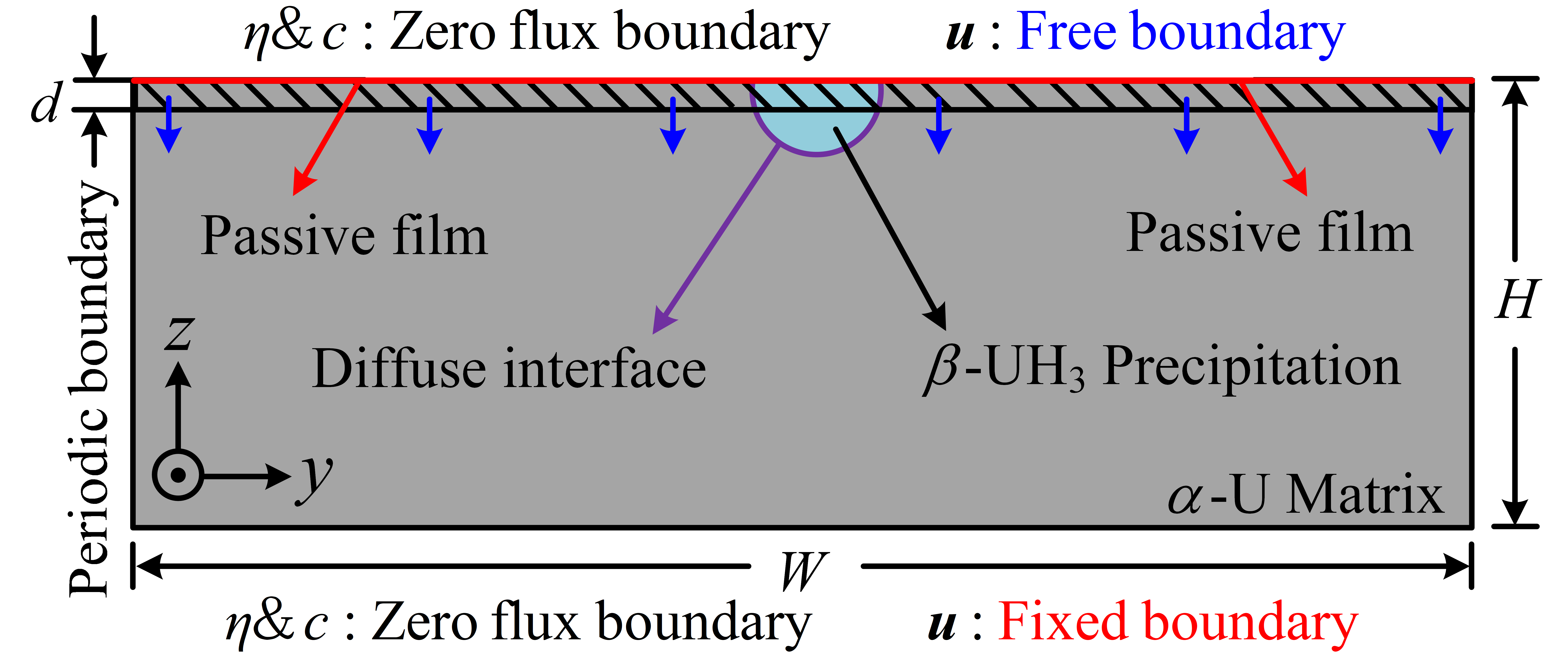}
	\caption{\label{fig2} Model diagram of the hydrogen-induced pitting corrosion in $\alpha$-U and boundary conditions setting. The $\alpha$-U matrix and $\beta$-UH$_{3}$ precipitate phases are separated by a diffuse interface. The shaded part is a hydrogen source used to simulate hydrogen from the environment entering the matrix. The blue arrows represents the hydrogen from the hydrogen source.}
\end{figure}

As shown in Fig. \ref{fig2}, an $\alpha$-U matrix with the initial $\beta$-UH$_{3}$ precipitate is regarded as a corrosion system studied. The width and height of the matrix are $W$ and $H$, respectively. For simplicity, it is assumed that the diffusion coefficients of hydrogen in the matrix and precipitate are the same. A hydrogen source region with the thickness of $d$ is set at the shaded part to simulate hydrogen from the environment entering the matrix\cite{guo2008elastoplastic}. The field variable or order parameter $\eta$ is used to distinguish between $\alpha$-U matrix ($\eta = 0$) and $\beta$-UH$_{3}$ precipitate ($\eta = 1$) phases. The field variable \textit{c} (atom fraction) is used to describe the hydrogen concentration variation of the system, and the mechanical interaction between the precipitate and the matrix is considered through the displacement field \textit{\textbf{u}}. Fig. \ref{fig2} also shows the boundary conditions for the simulated region. For both the phase and concentration field, the top and bottom boundaries are set to zero flux boundary conditions. For the displacement field, the top boundary is set free, while the bottom boundary is constrained. In addition, the periodic boundary conditions for all dependent variables are applied to the left and right boundaries\cite{abubakar2015phase}. 

In our model, the oxide film is considered to be the top boundary of zero thickness. It is a reasonable approximation in our studying for the facts as follow: (1) The typical thickness of the oxide film is only tens of nanometers\cite{harker2013altering} compared to the micron-sized hydrides; (2) The zero flux top boundary could act as a barrier between hydrogen and hydride, which is similar to the role of oxide film in the actual system; (3) It is believed that the oxide film may not play an important role in the pit morphology of the \uppercase\expandafter{\romannumeral2} period\cite{stitt2015effects}. A similar simplification also is carried out in the works of Mai \textit{et al.}\cite{mai2016phase,mai2017phase}. In addition, the ability of hydrogen passing through the oxide film could be simulated by adjusting the size and value of the hydrogen source. In other words, our model only considers the barrier effect of oxide film on hydrogen diffusion and ignores the mechanical properties of oxide film. Because the influence of mechanical properties of oxide film on pitting corrosion needs to be established under a theoretical framework of complex plastic large deformation. We will consider these effects in the future.

\subsection{PF governing equation}
In the proposed PF model, the system's free energy includes the bulk free energy $F_\text{bulk}$, the interface gradient energy $F_\text{int}$, and the elastic strain energy $F_\text{el}$, which is given by:
\begin{equation}
	\begin{aligned}
		\label{eq1}
		F& = F_\text{bulk} + F_\text{int} + F_\text{el} \\ 
		&= \int{[f_\text{bulk}(c,\eta) + f_\text{int}(\nabla \eta,\nabla c)]}dV + F_\text{el}(\textbf{\textit{u}},\eta).
	\end{aligned}
\end{equation}
$f_\text{bulk}(c,\eta)$ is the bulk free energy density and  $f_\text{int}(\nabla \eta,\nabla c)$ is the gradient energy density due to the diffuse interface, and the elastic strain energy is a function of displacement \textbf{\textit{u}} and order parameter $\eta$. In the PF frame, the gradient energy density could be written as a function of the gradient of the field variable:
\begin{equation}
	\begin{aligned}
		\label{eq2}
		& f_\text{int}(\nabla \eta,\nabla c)= \frac{1}{2}\kappa_{\eta}(\nabla\eta)^2 + \frac{1}{2}\kappa_{c}(\nabla c)^2, 		
	\end{aligned}
\end{equation}
where $\kappa_{\eta}$ and $\kappa_{c}$ are the gradient energy coefficients associated with the phase and concentration fields, respectively. Due to the fact that only one of the gradient terms ($\nabla c$  or $\nabla\eta$) could be sufficient to approximate the energy contribution from the diffuse interface\cite{mai2016phase}, we ignore the contribution of concentration gradient energy (assuming $\kappa_{c}$ = 0).

The PF governing equation followed by the evolution of the system could be derived by minimizing the total free energy \textit{F} via variational differentiation as follows:
\begin{align} 
		&\frac{\partial \eta}{\partial t} = -L\frac{\delta F}{\delta \eta} = -L\left(\frac{\partial f_\text{bulk}}{\partial \eta} - \kappa_{\eta}\nabla^2\eta+\frac{\delta F_\text{el}}{\delta \eta}\right),\label{eq3}\\ 		
		&\frac{\partial c}{\partial t} = \nabla\cdot \left( M\nabla\frac{\delta F}{\delta c}\right)  + S = \nabla\cdot \left( M\nabla\frac{\partial f_\text{bulk}}{\partial c}\right)  + S,\label{eq4}
\end{align}
where $L$ is the kinetic coefficient and $M$ is the mobility. The definition of the value of \textit{M} is similar to one used by Steinbach and Apel as follows\cite{steinbach2006multi}: $M=D/(\partial^2 f_\text{bulk}/\partial c^2)$, where $D$ is the diffusion coefficient\cite{powell1973mass}. In addition, a source term $S$ is added to Eq. \eqref{eq4} to simulate hydrogen from the environment entering the matrix\cite{guo2008elastoplastic}. This source term only exists at the shaded part in Fig. \ref{fig2}. Eqs. \eqref{eq3} and \eqref{eq4} respectively show that the evolution of order parameter $\eta$ and hydrogen concentration $c$ in time and space obey Ginzburg-Landau (also known as Allen-Cahn) and Cahn-Hilliard equations\cite{ansari2018phase}.  

\subsection{Bulk free energy density}
The bulk free energy density could be constructed by the KKS model\cite{kim1999phase}, and the model parameters could be analytically determined by material properties and experimental conditions\cite{ansari2018phase}. In the KKS model, each material point is regarded as a mixture of two phases, and a local equilibrium of chemical potential is always satisfied between the two phases. In this PF model, these assumptions could be expressed as follows:
\begin{align}
	&c = [1-h(\eta)]c_{\alpha} + h(\eta)c_{\beta},\label{eq5}\\
	&\frac{\partial f_{\alpha}(c_{\alpha})}{\partial c_{\alpha}} = \frac{\partial f_{\beta}(c_{\beta})}{\partial c_{\beta}},\label{eq6}
\end{align}
where $c_{\alpha}$ and $c_{\beta}$ represent the mole fraction of hydrogen in the $\alpha$-U matrix and $\beta$-UH$_{3}$ precipitate respectively, and $h(\eta)$ is a monotonously increasing interpolation function from $h(0)=0$ to $h(1)=1$. In this work, it is assumed that $h(\eta)=3\eta^2-2\eta^3$. In addition, $f_{\alpha}(c_{\alpha})$ and $f_{\beta}(c_{\beta})$  in Eq. \eqref{eq6} are the free energy densities of $\alpha$-U and $\beta$-UH$_{3}$ respectively. The bulk free energy density could be expressed by a method similar to the concentration assumption in Eq. \eqref{eq5} as follows:
\begin{equation}
	\begin{aligned}
		\label{eq7}
		&f_\text{bulk}(c,\eta)=[1-h(\eta)]f_{\alpha}(c_{\alpha})+h(\eta)f_{\beta}(c_{\beta})+wg(\eta),		
	\end{aligned}
\end{equation}
where $w$ is the height of the double potential well $g(\eta) = \eta^2(1-\eta)^2$. Eq. \eqref{eq7} in $\eta = 0$ and $\eta = 1$ has two minima, representing the $\alpha$-U matrix and $\beta$-UH$_{3}$ precipitate phases respectively.

The free energy density $f_{\alpha}(c_{\alpha})$ and $f_{\beta}(c_{\beta})$ could be constructed based on the parabolic approximation\cite{yang2020multiphase} similar to Bair \textit{et al.}'s work\cite{bair2017formation}. It is assumed that the free energy density $f_{\alpha}(c_{\alpha})$ and $f_{\beta}(c_{\beta})$ could be written as: 
	\begin{align}
		&f_{\alpha}(c_{\alpha})=A_{\alpha}(c_{\alpha}-c_{1})^2+f_{1},\label{eq8}\\
		&f_{\beta}(c_{\beta})=A_{\beta}(c_{\beta}-(c_{2}+\delta c))^2+f_{2},\label{eq9}\\
		&f_{2}=\Delta G/V_{m}, \label{eq10} \\
		&\Delta G = 0.1816T-127.67 \rm \left[kJ/mol\right], \label{eq11}
	\end{align}
where $c_{1}$ is the concentration of maximum hydrogen solubility in the matrix before precipitate occurs\cite{morrell2013uranium}; $f_{1}$ is the decreasing free energy caused by hydrogen dissolution in the matrix. Since the solubility of hydrogen in the $\alpha$-U matrix is very small, this item could be ignored, namely, $f_{1} \approx 0$\cite{bair2017formation}. Furthermore, $c_{2}$ is the concentration of hydrogen in the $\beta$-UH$_{3}$ precipitate; $f_{2}$ is the free energy of formation per unit volume of $\beta$-UH$_{3}$, where $V_{m}$ is the molar volume of the system\cite{banos2018review} and $\Delta G$ is the molar Gibbs free energy of formation of $\beta$-UH$_{3}$\cite{chiotti1980hu}. The $\delta c$ is a small parameter that regulates the magnitude of the thermodynamic driving force\cite{bair2017formation}. The constants $A_{\alpha}$ and $A_{\beta}$ in Eqs. \eqref{eq8} and \eqref{eq9} control the bulk contribution to the interfacial energy and the tangents between the phases. These constants could be determined to construct the common tangent between the free energy density curves of the $\beta$-UH$_{3}$ precipitate and the $\alpha$-U matrix\cite{bair2017formation}.
\begin{figure}
	\centering
	\includegraphics[width=0.4\textwidth]{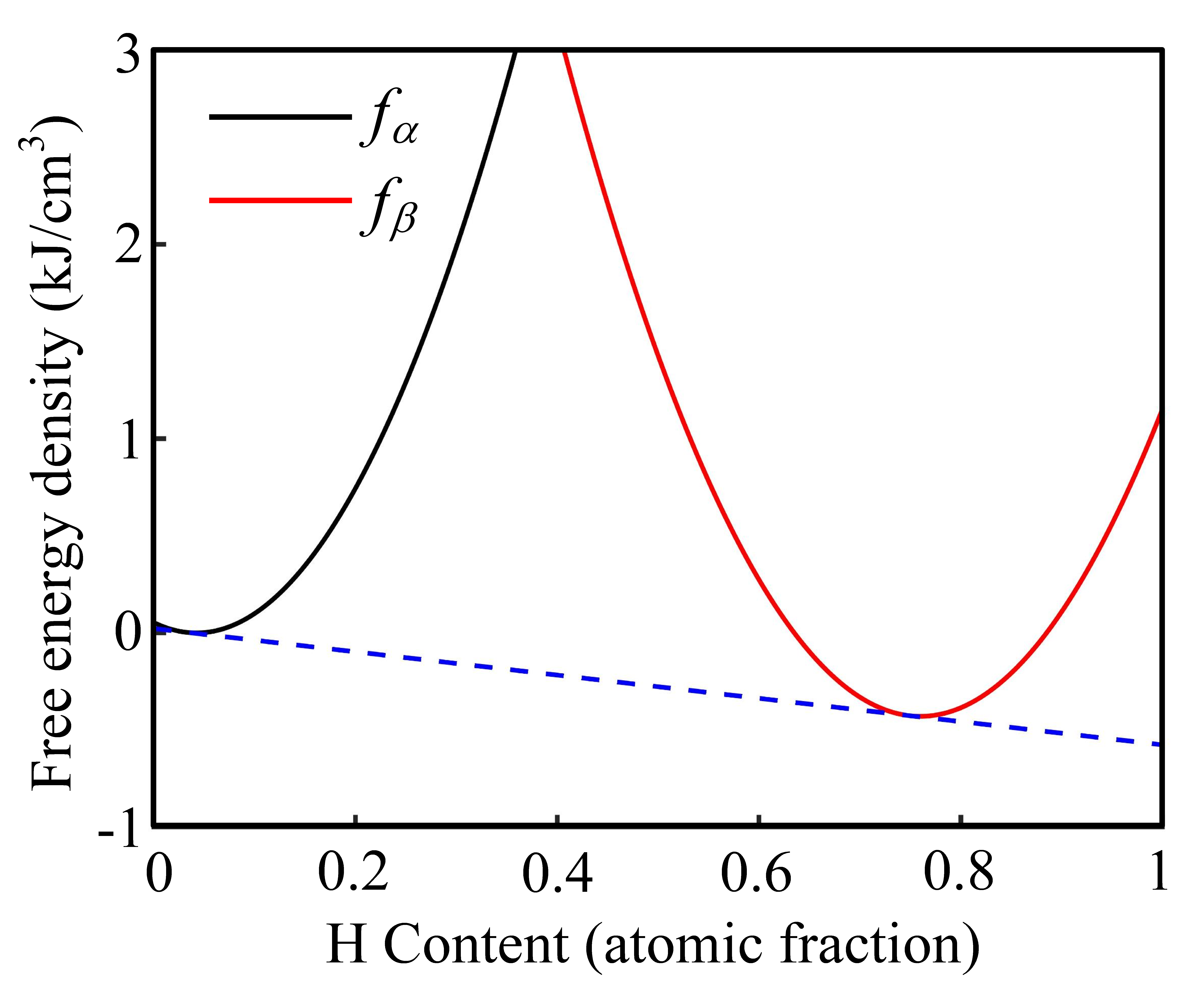}
	\caption{\label{fig3}Free energy densitiy curves of $\alpha$-U (black solid line, $f_{\alpha}$) and $\beta$-UH$_{3}$ (red solid line, $f_{\beta}$) in this work. The blue dotted line represents the phase equilibrium common tangent.}
\end{figure}
Fig. \ref{fig3} shows the free energy density curves $f_{\alpha}(c_{\alpha})$ and $f_{\beta}(c_{\beta})$ constructed in this work, where the black solid line represents  $f_{\alpha}(c_{\alpha})$, the red solid line represents $f_{\beta}(c_{\beta})$, and the common tangent line is represented by the dotted blue line. The detailed parameters of the free energy density curves can be found in Table \ref{tab1}.  

\begin{table}[]
	\caption{\label{tab1}Model parameters used in the simulations (eigenstrain $\varepsilon_{ij}^{00}$ is from the section \uppercase\expandafter{\romannumeral3} of supplementary materials ).}
	\begin{tabular}{cccc}
		\hline\hline
		Symbol    & Value                                               & Symbol                 & Value                                 \\
	    \hline
	     $T$      & 673 K                                               & $C_{11}^{\alpha}$      & 198.4 GPa\cite{fisher1966temperature} \\                                
	     $W$      & 2.5 $\upmu$m                                        & $C_{22}^{\alpha}$	     & 176.0 GPa\cite{fisher1966temperature} \\
	     $H$      & 0.5 $\upmu$m                                        & $C_{33}^{\alpha}$	     & 224.9 GPa\cite{fisher1966temperature} \\
	     $d$      & 50 nm                                               & $C_{12}^{\alpha}$	     & 53.7 GPa\cite{fisher1966temperature}  \\
	     $S$      & 0.05                                                & $C_{13}^{\alpha}$	     & 28.2 GPa\cite{fisher1966temperature}  \\
		 $A_{\alpha}$ & 29.51 kJ/cm$^{3}$                                     & $C_{23}^{\alpha}$	     & 98.7 GPa\cite{fisher1966temperature}  \\
		 $A_{\beta}$  & 27.08   kJ/cm$^{3}$           	                    & $C_{44}^{\alpha}$      & 54.6 GPa\cite{fisher1966temperature}  \\   
	   	 $c_{1}$      & 3.23 at.\%\cite{morrell2013uranium}             & $C_{55}^{\alpha}$	     & 95.8 GPa\cite{fisher1966temperature}  \\                     
		 $c_{2}$      & 75 at.\%                                        & $C_{66}^{\alpha}$      & 42.6 GPa\cite{fisher1966temperature}  \\
		 $V_{m}$      & 12.64 cm$^{3}$/mol\cite{banos2018review}        & $C_{11}^{\beta}$       & 227 GPa\cite{taylor2009ab}            \\
	     $\delta c$   & 0.01                                            & $C_{12}^{\beta}$ 	     & 102 GPa\cite{taylor2009ab}            \\
		 $L$          & 3.24e-9 m$^{3}$/(J·s)                           & $C_{44}^{\beta}$ 	     & 60  GPa\cite{taylor2009ab}            \\
		 $D$          & 0.019{\rm exp}(-5840/\textit{T}) cm$^{2}$/s \cite{powell1973mass}       & $\varepsilon^{00}_{11}$  &  0.5579 \\
		 $\lambda$    & 45 nm                                           & $\varepsilon^{00}_{22}$  &  0.0111  \\
		 $\chi$  & 20 J/m$^{2}$                                       & $\varepsilon^{00}_{33}$  &  0.1964      \\
	  	 $\kappa_{\eta}$  & 6.75e-7 J/m                                 & \multicolumn{1}{c}{\multirow{2}{*}{$\varepsilon^{00}_{ij}(i\neq j)$}}  & \multicolumn{1}{c}{\multirow{2}{*}{0}}         \\
		 $w$              & 1.07e10 J/m$^{3}$                           & \multicolumn{1}{c}{}  & \multicolumn{1}{c}{}         \\
         \hline\hline
	\end{tabular}	
\end{table}

\subsection{Elastic strain energy}
In this model, the mechanical interaction between metal and hydride is considered through introducing elastic strain energy, which is expressed as:

\begin{equation}
	\begin{aligned}
		\label{eq12}
		&F_\text{el} =\int f_\text{el}dV=\frac{1}{2} \int_{V}\sigma_{ij}\varepsilon_{ij}^{\mathrm{el}} d\textbf{r}, 		
	\end{aligned}
\end{equation}
where $f_\text{el}$ is the elastic strain energy density, and $\sigma_{ij}$ and $\varepsilon_{ij}^{\mathrm{el}}$ are stress and elastic strain respectively. Einstein summation convention is used here. 
By using the additive decomposition theorem of strain under the assumption of small strain and ignoring the plastic strain, the total strain could be written as the sum of elastic strain and stress-free strain, so the elastic strain is given by:
\begin{equation}
	\begin{aligned}
		\label{eq13}
		&\varepsilon_{ij}^{\mathrm{el}} = \varepsilon_{ij}^{\mathrm{tot}} - \varepsilon_{ij}^{\mathrm{sf}}, 		
	\end{aligned}
\end{equation}
where $\varepsilon_{ij}^{\mathrm{tot}}$ is the total strain, and $\varepsilon_{ij}^{\mathrm{sf}}$ is the stress-free strain.

Using the classical Khachaturyan micro-elasticity theory\cite{morris2010khachaturyan}, the stress-free strain is expressed by:
\begin{equation}
	\begin{aligned}
		\label{eq14}
		&\varepsilon_{ij}^{\mathrm{sf}} = h(\eta)\varepsilon_{ij}^{00} = h(\eta)\begin{bmatrix} 
			\varepsilon_{11}^{00}&\varepsilon_{12}^{00}&\varepsilon_{13}^{00} 
			\\ \varepsilon_{21}^{00}&\varepsilon_{22}^{00}&\varepsilon_{23}^{00} 
			\\ \varepsilon_{31}^{00}&\varepsilon_{32}^{00}&\varepsilon_{33}^{00}      
		    \end{bmatrix},  		
	\end{aligned}
\end{equation}
where $\varepsilon_{ij}^{00}$ is the eigenstrain, depicting the degree of lattice mismatch and volume expansion during the phase transformation. Eq. \eqref{eq14} means that eigenstrain $\varepsilon_{ij}^{00}$ only exists in the $\beta$-UH$_{3}$ precipitate\cite{yeon2005phase}. 

With Lagrangian finite-strain theory\cite{lubliner2008plasticity},  the eigenstrain from $\alpha$-U to $\beta$-UH$_{3}$ is expressed as:
\begin{equation}
\begin{aligned}
	\label{eq15}
	\varepsilon_{ij}^{00}=\frac{1}{2}(U^TU-I)
\end{aligned}
\end{equation}
where $U$ is the total lattice transformation tensor or deformation gradient tensor\cite{han2019phase} from $\alpha$-U to $\beta$-UH$_{3}$ (the calculation method of $U$ is shown in the supplementary materials). The superscript $T$ represents the matrix transpose and $I$ represents the identity matrix.

According to the micro-elasticity theory\cite{morris2010khachaturyan}, the total strain $\varepsilon_{ij}^{\mathrm{tot}}$ is the sum of macroscopic homogeneous strain and microscopic heterogeneous strain, which is given by:
\begin{align}
	&\varepsilon_{ij}^{\mathrm{tot}} = \bar{\varepsilon}_{ij}+\delta\varepsilon_{ij},\label{eq16}\\
	&\bar{\varepsilon}_{ij} = \frac{1}{V}\int\varepsilon_{ij}^{\mathrm{sf}}dV,\label{eq17}\\
	&\delta\varepsilon_{ij}=\frac{1}{2}\left(\frac{\partial u_i}{\partial r_j}+\frac{\partial u_j}{\partial r_i}\right),\label{eq18}
\end{align}
where $\bar{\varepsilon}_{ij}$ is the macroscopic homogeneous strain, $\delta\varepsilon_{ij}$ is the microscopic heterogeneous strain. The macroscopic homogeneous strain represents the variation of the macroscopic shape of the system in the corrosion process. It could be calculated by minimizing the elastic strain energy relative to the macroscopic homogeneous strain, and equals to the volume average value\cite{zhang2005phase} of the stress-free strain $\varepsilon_{ij}^{\mathrm{sf}}$, as shown in Eq. \eqref{eq17}. The microscopic heterogeneous strain is related to the displacement, as shown in Eq. \eqref{eq18}. The displacement is obtained by solving the mechanical equilibrium equation as follows:
\begin{equation}
	\begin{aligned}
		\label{eq19}
		&\frac{\partial \sigma_{ij}}{\partial r_j} = 0.
	\end{aligned}
\end{equation}
Here, we assume that both the $\alpha$-U matrix and the $\beta$-UH$_{3}$ precipitate obey linear elastic theory, so the relationship between stress and elastic strain is as follows: 
\begin{align}
	&\sigma_{ij} = C_{ijkl}(\eta)\varepsilon_{kl}^{\mathrm{el}},\label{eq20}\\
	&C_{ijkl}(\eta) = [1-h(\eta)]C_{ijkl}^{\alpha}+h(\eta)C_{ijkl}^{\beta},\label{eq21}
\end{align}
where $C_{ijkl}(\eta)$ is the global elastic constant\cite{yeon2005phase} with associated to the order parameter $\eta$, $C_{ijkl}^{\alpha}$ and $C_{ijkl}^{\beta}$ are the elastic constants of the $\alpha$-U matrix\cite{fisher1966temperature} and the $\beta$-UH$_{3}$ precipitate\cite{taylor2009ab} respectively.

\subsection{Numerical implementation}
In the PF frame, the gradient energy coefficient $\kappa_{\eta}$ and the double-well height $w$ are correlated to the interface energy $\chi$ and the interface thickness $\lambda$ as follows\cite{kim1999phase}:  
\begin{align}
	&\chi = \frac{\sqrt{\kappa_{\eta}w}}{3\sqrt{2}},\label{eq22}\\
	&\lambda = \xi\sqrt{\frac{2\kappa_{\eta}}{w}},\label{eq23}
\end{align}
where $\xi = 4$ is a constant parameter associated with the definition of the interface region, and the interface region is defined as $0.018<\eta<0.982$\cite{yang2021explicit} in this work. The interface thickness selected in the current PF model is significantly smaller than the characteristic length scale of the problem\cite{mai2016phase}.
\begin{figure*}[ht]
	\centering
	\includegraphics[width=0.95\textwidth]{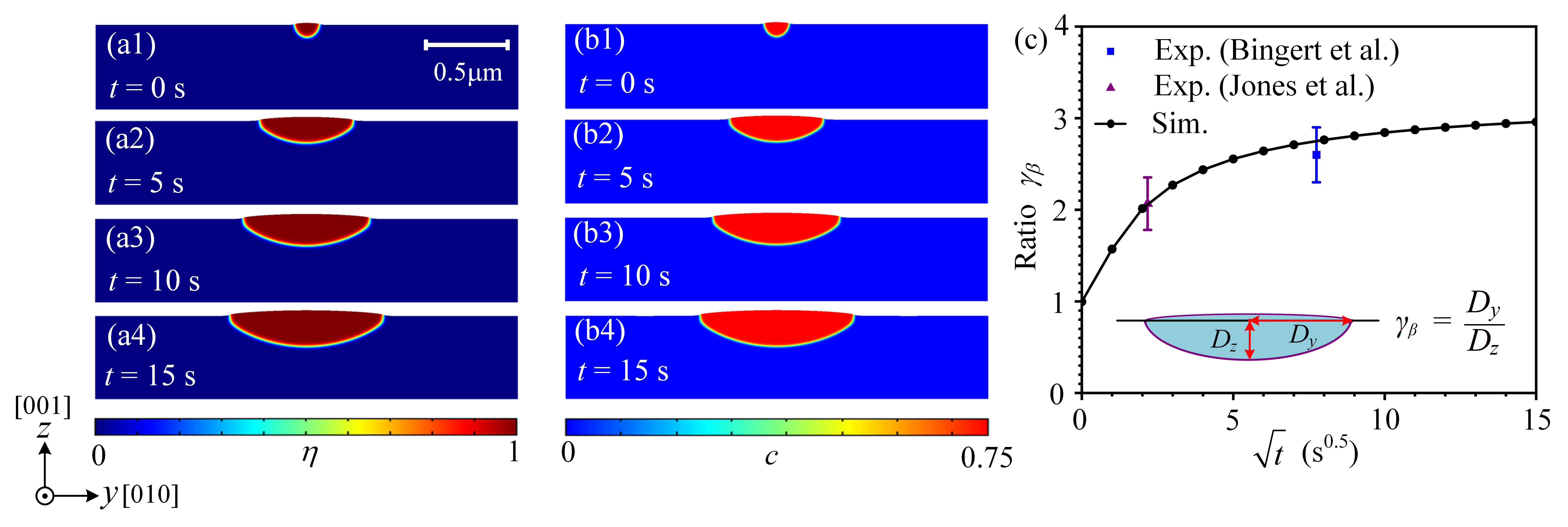}
	\caption{\label{fig4}(a1)-(a4) Evolution of the pit morphology in the $\alpha$-U matrix at different times $t$. (b1)-(b4) distribution of hydrogen concentration $c$ during the pitting process of $\alpha$-U matrix at different times $t$. (c) variation of the shape factor $\gamma_{\beta}$ versus $\sqrt{t}$ in the simulation, blue and purple dots represent the calculated shape factors based on the experimental observations from Refs.~\onlinecite{bingert2004microtextural} and \onlinecite{jones2013surface}, respectively. Inset shows the definition of shape factor.}
\end{figure*}

In order to solve the evolution equation, the FEM is adopted  \cite{mai2016phase,ansari2018phase,bair2016phase,abubakar2015phase}. The reason for choosing FEM is that the system studied in this work has non-periodic boundaries, which is difficult to be solved by traditional periodic Fourier spectral method\cite{li2017review,yang2021hydride}. To ensure the stability and accuracy of numerical simulation, the 6-node triangular Lagrange element is used to approximate the governing equation, and the thickness of the interface region contains at least 18 elements to properly approximate $\eta$. To avoid using an excessively refined finite element mesh everywhere, we use the magnitude of $\left| \nabla\eta\right|$ as a refinement criterion\cite{mai2016phase} to implement a local adaptive mesh refinement scheme to provide higher mesh density at the interface region where it is most needed. The initial Lagrange-quadratic elements of the domain are about 17,000. Moreover, an implicit backward difference method and an adaptive time marching scheme with a tolerance of $10^{-3}$ were adopted to assure the numerical convergence, while keeping the computational efficiency. 

For the concrete simulation, we adopt a similar approach as Mai \textit{et al.}'s work\cite{mai2016phase}. The simulation is carried out under the plane stress assumption. At the beginning of the simulation, a semi-circular initial $\beta$-UH$_{3}$ precipitate with the radius of 50 nm is placed at the top of the rectangular simulation area. The hydrogen concentration of initial precipitate is set to 0.75, and that of $\alpha$-U matrixis set to 0.05 for providing a large thermodynamic driving force for hydrogen-induced pitting corrosion\cite{han2019phase}. Unless otherwise stated, values of the parameters used for performing the simulations presented in the following sections are selected from Table \ref{tab1}.

In our work, we set [100], [010], [001]-direction of $\alpha$-U along the $x, y, z$-axis of the simulated area (see section \uppercase\expandafter{\romannumeral2} of supplementary materials). The (001) surface of $\alpha$-U is exposed to the hydrogen, since some studies have indicated that this surface is prone to hydrogen absorption and hydrogen-induced pitting corrosion\cite{taylor2009ab,taylor2008evaluation}.  
The plane used in our simulation is (100) surface.

\section{Results and Discussion}\label{sec3} 

\subsection{Morphology of pitting growth}

We use this proposed PF model to study the pit morphology and the distribution of hydrogen during the pitting process of $\alpha$-U matrix in this subsection. Figs. \ref{fig4}(a1)-\ref{fig4}(a4) show the evolution of the pit morphology during the pitting process of $\alpha$-U matrix at different times. $\eta$ is defined as an order parameter desribing phases in section \ref{sec2}. Thus, the red part ($\eta = 1$) represents the hydride precipitate phase in Figs. \ref{fig4}(a1)-(a4), while the blue part ($\eta = 0$) represents the $\alpha$-U matrix phase. The remaining part represents a diffuse interface separating two phases. Figs. \ref{fig4}(b1)-\ref{fig4}(b4) show the hydrogen concentration $c$ in the hydride precipitate, $\alpha$-U matrix and diffuse interface at different times. To verify the accuracy of the model, we compared the pit morphology of simulation with the ones of experiments presented in Refs.~\onlinecite{bingert2004microtextural} and \onlinecite{jones2013surface}, as shown in Fig. \ref{fig4}(c). The black solid line represents the simulations, and the blue and purple dots represent the calculated shape factors based on the experimental observations from Refs.~\onlinecite{bingert2004microtextural} and \onlinecite{jones2013surface}, respectively. It should be noted that Jones \textit{et al.}'s work\cite{jones2013surface} did not give an exact time for pitting corrosion, so the abscissa ($\sqrt{t}$) of this data point was estimated by the simulation. 
   
Both Figs. \ref{fig4}(a1)-\ref{fig4}(a4) and Figs. \ref{fig4}(b1)-\ref{fig4}(b4) illustrate that the pit morphology or hydride precipitate is anisotropic, and approximates an ellipse with $y$-direction as its major axis and $z$-direction as its minor axis. In addition, the other feature is that a slight bulge appears on the top boundary of $\alpha$-U matrix from the these Figures. These results suggest that our PF model has the ability to simulate an anisotropic pit morphology and the bulge on the metal surface in hydrogen-induced pitting corrosion.

For the quantitative comparison of pit morphology in Fig. \ref{fig4}(c), we define a shape factor $\gamma_{\beta}$ of hydride precipitate as follows: 
\begin{align}
	&\gamma_{\beta} =\frac{D_{y}}{D_{z}},\label{eq24}
\end{align}
where $D_{y}$ and $D_{z}$ are pit depth of $y$-direction and $z$-direction, respectively. Fig. \ref{fig4}(c) illustrates that the shape factor increases gradually with time, it implies that the ellipticity of pit morphology increases gradually with time. Moreover, the simulation is in good agreement with the experiments\cite{jones2013surface,bingert2004microtextural}. It suggests that this proposed PF model is validity.

In our opinion, the anisotropic pit morphology could be simulated in our PF model results from that the minimization of free energy including elastic strain energy is considered in this PF method. In actual corrosion system, the pit morphology could be estimated by the tendency to follow the path of lowest system's free energy\cite{greenbaum2008strain}. Moreover, the elastic strain energy has the most significant effect on pit morphology in the system's free energy\cite{han2019phase}. This proposed PF model is based on these two important physical laws.

\begin{figure*}[ht]
	\centering
	\includegraphics[width=1\textwidth]{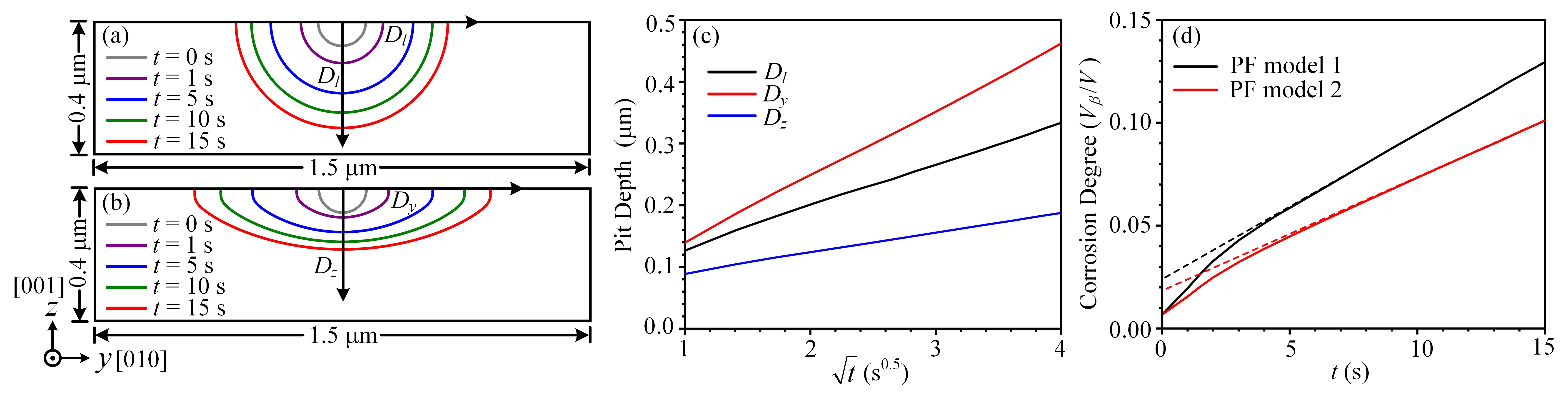}
	\caption{\label{fig5}(a) Propagation of the pitting corrosion interface (metal-hydride interface) simulated via the PF model without coupling elastic strain energy in the Lagrangian coordinates; (b) propagation of the pitting corrosion interface simulated via our PF model in the Lagrangian coordinates; (c) variation of the pit depth of both PF models versus $\sqrt{t}$; (d) variation of the corrosion degree ($V_{\beta}/V$) of both PF models versus $t$, PF model 1: PF model without coupling elastic strain energy, PF model 2: our PF model.}
\end{figure*}

\subsection{Kinetics of pitting growth}
To investigate the kinetics of pitting growth, we compare the results of our PF model and PF model without coupling elastic strain energy\cite{mai2016phase}. Figs. \ref{fig5}(a) and (b) show the propagation of pitting corrosion interface versus time $t$ using the PF model without coupling elastic strain energy and our PF model, respectively. $D_{l}$ represents the pit depth simulated via the PF model without coupling elastic strain energy. $D_{y}$ and $D_{z}$ represents the pit depth of $y$-direction and $z$-direction in our PF model. Fig. \ref{fig5}(a) illustrates an isotropic pitting process. The pitting kinetics given by our PF model is anisotropic in Fig. \ref{fig5}(b), and the corrosion of $y$-direction is faster than the one of $z$-direction. The results of Fig. \ref{fig5}(a) and \ref{fig5}(b) also confirm that our model has the ability to simulate an anisotropic pitting kinetics. The anisotropy is enhanced during the pitting process in our PF model, while the result from PF model without coupling elastic strain energy still keep an isotropic feature.

Fig. \ref{fig5}(c) shows the variations of pit depth versus $\sqrt{t}$. It illustrates that $D_{y}>D_{l}>D_{z}$, and the pit depth of in both PF models is approximately proportional to $\sqrt{t}$. These results imply that the pitting growth conforms to an anisotropic parabolic pitting kinetics dominated by diffusion control\cite{mai2016phase}. These are consistent with a fact that the diffusion of hydrogen is the main limiting factor in the formation of hydride precipitates\cite{bair2017formation}. 
 
We also study the variations of corrosion degree ($V_{\beta}/V$) simulated via both PF models versus $t$, as shown in Fig. \ref{fig5}(d). The $V_{\beta}$ represents the volum of the hydride precipitate, and $V$ is total volum of system. Fig. \ref{fig5}(d) illustrates that the corrosion degree is approximately proportional to time. The result simulated by the PF model without coupling elastic strain energy is larger than that simulated by our PF model, and overestimate the degree of pitting corrosion. The part where the dotted line deviates from the solid line in Fig. \ref{fig5}(d) may be influenced by the initial nucleation size.

From the results simulated via both PF models, our PF model could describe the anisotropic pitting kinetics with solid-solid phase transformation, such as hydrogen-induced pitting corrosion, but the common PF model without coupling elastic strain energy could not. The common PF model is more suitable for pitting corrosion with solid-liquid phase transformation, such as the pitting corrosion of the metal in the electrolyte\cite{mai2016phase}. Because the elastic constant of the liquid phase and the stress-free strain related to the solid-liquid phase transformation are almost zero, the elastic strain energy of the system could be ignored without applied stress or strain. Thus, the pit morphology and kinetics are prone to perform an isotropic feature in the metal-electrolyte system\cite{mai2016phase}.

\begin{figure*}[ht]
	\centering
	\includegraphics[width=0.85\textwidth]{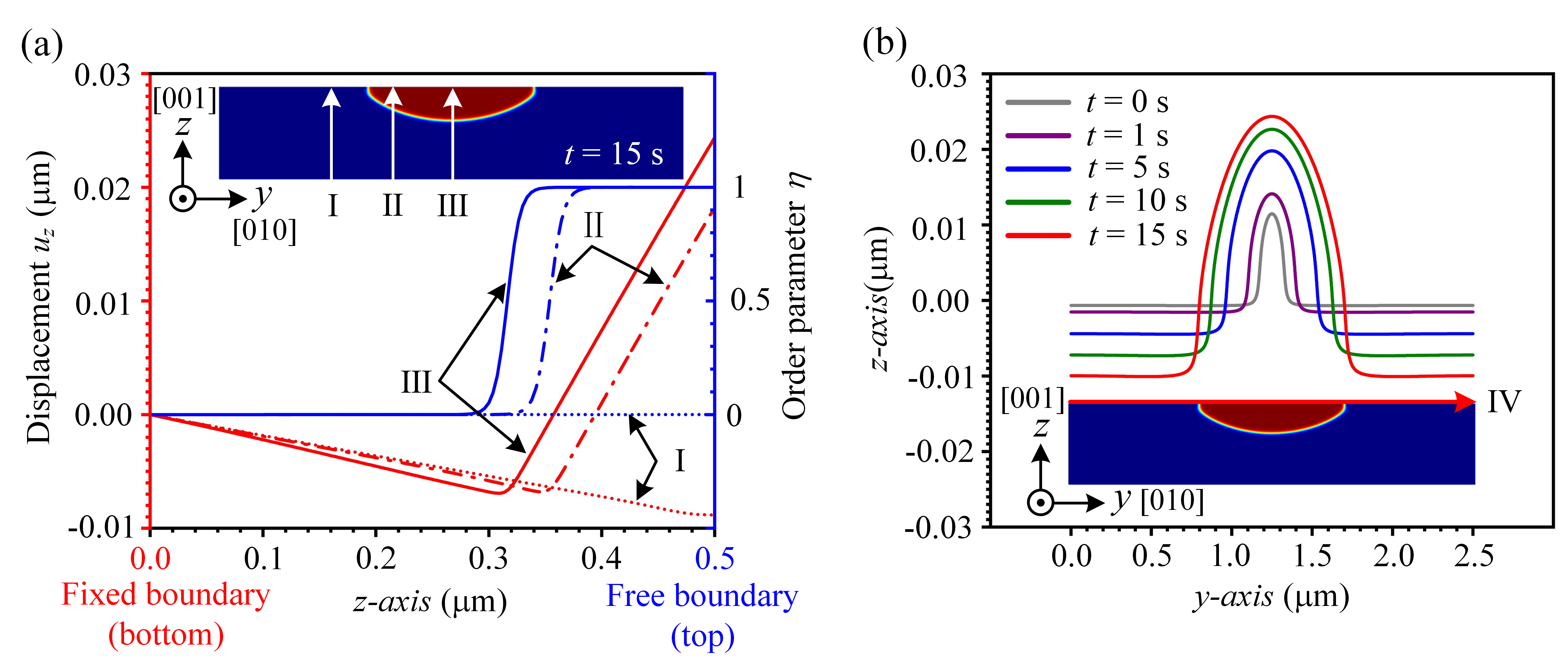}
	\caption{\label{fig6}(a) Variation of the displacement component $u_{z}$ (red line) and order parameter $\eta$ (blue line) along lines \uppercase\expandafter{\romannumeral1}, \uppercase\expandafter{\romannumeral2} and \uppercase\expandafter{\romannumeral3} at $t = 15$ s; (b) deformation of line \uppercase\expandafter{\romannumeral4} (i.e. top boundary) at different times. The pit morphology of the insets of (a) and (b) are in the Lagrange coordinates.}
\end{figure*}
\subsection{Kinematics analysis}
We investigate the displacement and deformation of material points in the pitting process of $\alpha$-U matrix via our PF model and analyze how displacement and deformation affect the pit morphology.

Fig. \ref{fig6}(a) displays the variation of the displacement component $u_{z}$ (red line) and order parameter $\eta$ (blue line) along the lines \uppercase\expandafter{\romannumeral1}, \uppercase\expandafter{\romannumeral2} and \uppercase\expandafter{\romannumeral3} of the inset. Line \uppercase\expandafter{\romannumeral1} only passes through the matrix, while line \uppercase\expandafter{\romannumeral2} and line \uppercase\expandafter{\romannumeral3} both pass through the matrix and hydride, and line \uppercase\expandafter{\romannumeral3} passes through more of the hydride. The displacement component $u_{z}$  represents the movement direction and distance of the material points, positive: toward the top boundary; negative: toward the bottom boundary.

The displacement component $u_{z}$  along the line \uppercase\expandafter{\romannumeral1} decreases from zero (due to the fixed bottom boundary) monotonously. It means that the matrix in this part are mainly compressed, and the closer they are to hydride precipitate, the stronger the compression effect is. However, the displacement components $u_{z}$ along lines \uppercase\expandafter{\romannumeral2} and \uppercase\expandafter{\romannumeral3} show different features from line \uppercase\expandafter{\romannumeral1} when crossing the hydride precipitate. The displacement components $u_{z}$ of the matrix points reach a minimum near the diffuse interface. After passing the diffuse interface, the displacement components $u_{z}$ begin to increase and soon appear a positive displacement. It means that the hydride points below the top boundary move upward. The upward movements cause the deformation of the free top boundary, producing a bulge. Fig. \ref{fig6}(b) shows the deformation of line \uppercase\expandafter{\romannumeral4} (top boundary) of the inset at different times. It clearly displays the bulge on the free top boundary at different times. From Fig. \ref{fig6}(b), the deformation of the bulge on the top boundary increased with time, which illustrates that the hydride precipitate is dilation. In addition, the deformation of matrix is towards the bottom boundary, which is consistent with those in Fig. \ref{fig6}(a).  

An important feature is illustrated from these results that the matrix is compressed and the hydride precipitate is dilation during the pitting process. The feature could be explained by two effects\cite{greenbaum2011elastic}: (1) the hydride dilation leads to compression of the surrounding matrix; (2) the hydride bulge makes the material points move upward, toward the free top boundary. The influences of both effects are well captured in Fig. \ref{fig6}(a) and \ref{fig6}(b). The matrix points along line \uppercase\expandafter{\romannumeral1} move toward the bottom boundary in Fig. \ref{fig6}(a) because of the first effect. Due to the second effect, the hydride points along line \uppercase\expandafter{\romannumeral2} and \uppercase\expandafter{\romannumeral3} move toward the top boundary passing through the diffuse interface, forming a bulge on the free top boundary in Fig. \ref{fig6}(b). It should be noted that both effects affect the displacement component $u_{z}$ of the hydride, and the second effect dominates.

\begin{figure*}[ht]
	\centering
	\includegraphics[width=0.8\textwidth]{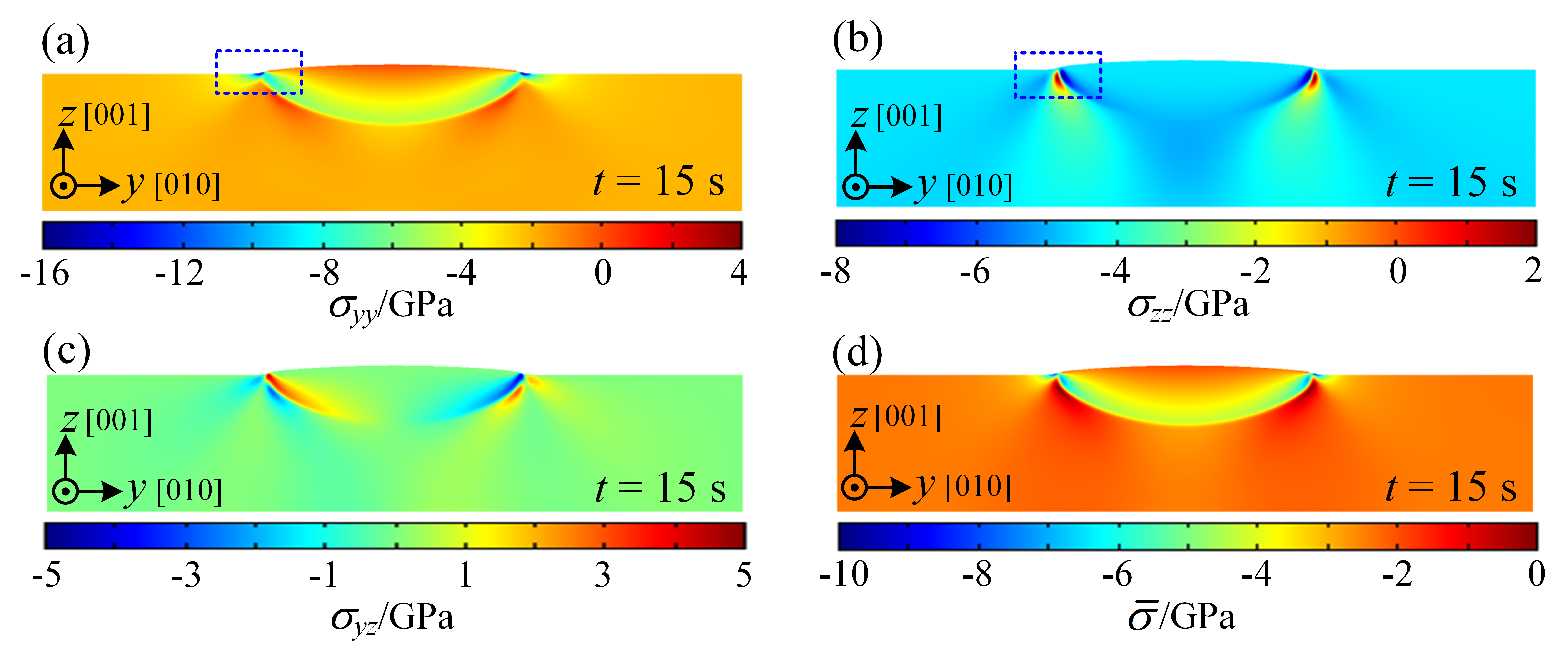}
	\caption{\label{fig7}Stress distribution of the corroded  $\alpha$-U matrix at $t = 15$ s. $>0$: tensile stress, $<0$: compressive stress. (a) $\sigma_{yy}$ and (b) $\sigma_{zz}$, the blue dotted boxes show the areas with stress concentration; (c) $\sigma_{yz}$; (d) $\bar{\sigma}$.}
\end{figure*}
\begin{figure*}[ht]
	\centering
	\includegraphics[width=1\textwidth]{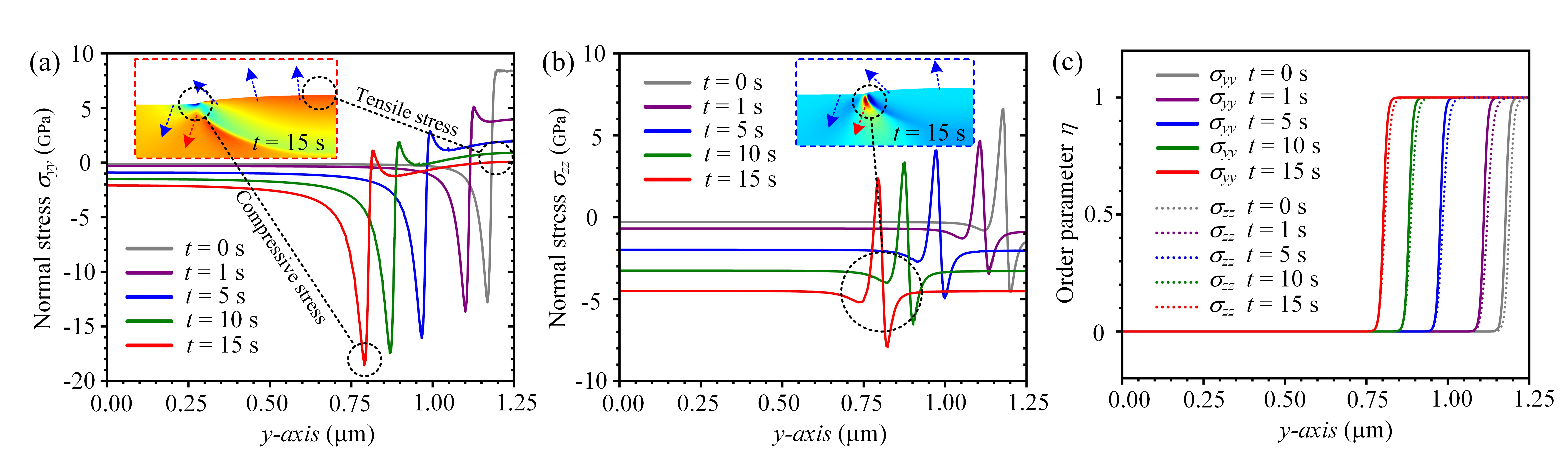}
	\caption{\label{fig8} Details of the stress concentration areas (blue dotted boxes of Fig. \ref{fig7}(a) and \ref{fig7}(b)), the blue and red dotted arrows represent the movement direction of the material points in the insets; $>0$: tensile stress; $<0$: compressive stress. (a) variation of $\sigma_{yy}$ along the $z = 0.5 \mu$m (top boundary); (b) variation of $\sigma_{zz}$ along the $z = 0.46 \mu$m; (c) variation of order parameter $\eta$ along the $z = 0.5 \mu$m (corresponding to $\sigma_{yy}$) and $z = 0.46 \mu$m (corresponding to $\sigma_{zz}$).}
\end{figure*}

\subsection{Stress analysis}
The last subsection focuses on the stress distribution during the pitting process and analyzes the influence of stress on the matrix and hydride. Figs. \ref{fig7}(a)-\ref{fig7}(d) show the stress distribution of normal stresses $\sigma_{yy}$, $\sigma_{zz}$, shear stress $\sigma_{yz}$ and hydrostatic stress $\bar{\sigma} = \frac{1}{3}(\sigma_{xx}+\sigma_{yy}+\sigma_{zz})$ at $t = 15$ s, respectively. Figs. \ref{fig7}(a) and \ref{fig7}(b) illustrate that the compressive stress govern most areas of both hydride precipitate and matrix.
There are some special areas (blue dotted boxes) with stress concentration near the ends of
the hydride precipitate in Figs. \ref{fig7}(a) and \ref{fig7}(b). Fig. \ref{fig7}(c) shows that the shear stress mainly exists near the matrix-precipitate interface. It is generally known that shear stress tends to cause deformation of a solid by slippage along a plane or planes parallel to the imposed stress. Therefore, Fig. \ref{fig7}(c) implies the deformation direction of the matrix-hydride interface. From the distribution of hydrostatic stress in Fig. \ref{fig7}(d), we could observe a possible dilational direction of the hydride precipitate\cite{guo2008elastoplastic,greenbaum2011elastic}.

Fig. \ref{fig8}(a) and \ref{fig8}(b) shows details of these areas with stress concentrations at different times. Fig. \ref{fig8}(c) displays the distribution of order parameter $\eta$ in these areas at different times. The blue and red dotted arrows represent the movement direction of the material points in the insets of Fig. \ref{fig8}(a) and (b). It can be seen that there is a compressive stress concentration of $\sigma_{yy}$ in Fig. \ref{fig8}(a). Near the compressive stress concentration, the movement direction of the material points change greatly, from stretching to compression. Fig. \ref{fig8}(b) and its inset display an area with stress concentration of $\sigma_{zz}$ near the end of the hydride precipitate, which is similar to the solution of the classical Eshelby's inclusion problem\cite{eshelby1957determination}. This stress concentration is the result of the interaction between hydride and matrix during the growth of hydride. In the growth of hydride, the ends of hydride tend to split the surrounding matrix, which makes the matrix is stretched around the ends of hydride. Meanwhile, the growth of hydride is blocked by the matrix, which makes the ends of hydride compressed. However, this local stretching of matrix does not prevent the overall compression of the matrix caused by hydride dilation, which can be seen from the displacement of the matrix point near this local stretching in inset of Fig. \ref{fig8}(b)(red arrow). It's worth noting that Fig. \ref{fig8}(a) also shows tensile stress near the hydride precipitate on the top boundary. The tensile stress decreases gradually over time, which may be related to the shape change of the growing hydride precipitation. In addition, another interesting feature is that the strength of stress concentrations (the sharpness of curves) in Figs. \ref{fig8}(a) and \ref{fig8}(b) over time varies very little.

Figs. \ref{fig7}(a) and \ref{fig7}(b) show that both the matrix and the hydride precipitate are mostly dominated by compressive stress. However, from the kinematics analysis in the section \ref{sec3} C, the matrix is slightly compressed and the hydride precipitate is dilation. It suggests that the strain causing the hydride dilation isn't induced by stress, implying a stress-free strain in the hydride. In our PF model, the stress-free strain is directly related to the eigenstrain describing the volume expansion during the phase transformation, as Eq. \ref{eq14}. Thus, the phase transformation is the reason for the hydride precipitate dilation.

The phenomenon that stress concentrates at both ends of the hydride precipitate in Fig. \ref{fig8} gives us some important insight into the hydride growth along the surface and the fracture during the pitting process. However, in the actual matrix, the surface of the matrix is usually covered with a thin oxide film, which is a porous and brittle ceramic material\cite{evans1969strength,harker2012corrosion}. Unfortunately, our current model only focuses on pit morphology and ignores the mechanical properties of the oxide film. To get a better understanding about the fracture during the pitting process, a pitting simulation explicitly considerating the mechanical properties of the oxide film may be needed.

\section{Conclusion}\label{sec4} 
In our work, we present a new PF model with elastic strain energy, and adopt a numerical technique with free boundary condition based on the FEM, and perform the simulation of hydrogen-induced pitting corrosion in $\alpha$-U with this PF model and numerical technique. For our PF model, the elastic strain energy is explicitly introduced to approximate the mechanical interaction between metal and hydride, which leads to a more practical anisotropic pit morphology. For the numerical technique, the free boundary condition based on the FEM is employed, which makes it possible to exhibit the bulges on the metal surface during the pitting process. In the simulation of hydrogen-induced pitting corrosion of $\alpha$-U, the pit morphology is approximately an elliptical shape, which is consistent with the experiments. Moreover, we further find that the formation of the bulge is caused by the compression of the metal and the dilation of the hydride, and it is attributed to the free boundary condition. Together with the analysis of stress of hydride, the hydride precipitate dilation is driven by the phase transformation. Another interesting finding is that the stress concentrates at both ends of the hydride precipitate, which is helpful for understanding the growth mechanism of hydride precipitate and the fracture during the pitting process. In addition, this PF model could be applied to other hydrogen-induced pitting corrosion systems, and the simulation analysis of $\alpha$-U provides ideas for corrosion protection in the nuclear industry.

\section{Acknowledgement}
We thank Hou-Bing Huang, Xiao-Ming Shi, Chao Yang, Chang-Qing Guo and Guo-Min Han for the help in the method of modeling. We thank Xing-Yu Gao, Dan Jian, Li-Fang Wang and Le Zhang for the help about basic knowledge and properties of uranium. We thank Yuan-Ji Xu, Bei-Lei Liu, Kai-Le Chen, Ji-De Zou for helpful discussions. The work was supported by the Science Challenge Project (NO. TZ2018002), the National Key R\&D Program of China (NO. 2021YFB3501503) and the Foundation of LCP. 

\section{Author Contributions}
H.-F. Song and Y. Liu conceived and supervised the project. J. Sheng performed the numerical simulations. All authors analysed and discussed the results. J. Sheng, Y.-C. Wang, Y. Liu, S. Wu and H.-F. Song wrote the manuscript, with contributions from all the authors.

\clearpage


\begin{thebibliography}{10}
	
	\bibitem{ernst2002pit1}
	P~Ernst and RC~Newman.
	\newblock Pit growth studies in stainless steel foils. \uppercase\expandafter{\romannumeral1}. introduction and pit
	growth kinetics.
	\newblock {\em Corrosion Science}, 44(5):927--941, 2002.
	
	\bibitem{ernst2002pit2}
	P~Ernst and RC~Newman.
	\newblock Pit growth studies in stainless steel foils. \uppercase\expandafter{\romannumeral2}. effect of
	temperature, chloride concentration and sulphate addition.
	\newblock {\em Corrosion Science}, 44(5):943--954, 2002.
	
	\bibitem{banos2018review}
	A~Banos, NJ~Harker, and TB~Scott.
	\newblock A review of uranium corrosion by hydrogen and the formation of
	uranium hydride.
	\newblock {\em Corrosion Science}, 136:129--147, 2018.

    \bibitem{alvarez2011phase}
    MAV Alvarez, JR~Santisteban, G~Domizzi, and J~Almer.
    \newblock Phase and texture analysis of a hydride blister in a Zr--2.5\% Nb
    tube by synchrotron X-ray diffraction.
    \newblock {\em Acta Materialia}, 59(5):2210--2220, 2011.	
	
	\bibitem{brierley2016anisotropic}
	M Brierley, JP Knowles, A Sherry, and M Preuss.
	\newblock The anisotropic growth morphology and microstructure of plutonium
	hydride reaction sites.
	\newblock {\em Journal of Nuclear Materials}, 469:145--152, 2016.
	
	\bibitem{bingert2004microtextural}
	JF~Bingert, RJ~Hanrahan~Jr, RD~Field, and PO~Dickerson.
	\newblock Microtextural investigation of hydrided $\alpha$-uranium.
	\newblock {\em Journal of Alloys and Compounds}, 365(1-2):138--148, 2004.
	
	\bibitem{jones2013surface}
	CP Jones, TB Scott, JR Petherbridge, and J Glascott.
	\newblock A surface science study of the initial stages of hydrogen corrosion
	on uranium metal and the role played by grain microstructure.
	\newblock {\em Solid State Ionics}, 231:81--86, 2013.
	
	\bibitem{ji2019mechanism}
	H Ji, H Wu, Q Pan, D Cai, X Meng, X Chen, P
	Shi, and X Wang.
	\newblock Mechanism of surface uranium hydride formation during corrosion of
	uranium.
	\newblock {\em npj Materials Degradation}, 3(1):1--8, 2019.
		
	\bibitem{scheiner2007stable}
	S Scheiner and C Hellmich.
	\newblock Stable pitting corrosion of stainless steel as diffusion-controlled
	dissolution process with a sharp moving electrode boundary.
	\newblock {\em Corrosion Science}, 49(2):319--346, 2007.
	
	\bibitem{scheiner2009finite}
	S Scheiner and C Hellmich.
	\newblock Finite volume model for diffusion-and activation-controlled pitting
	corrosion of stainless steel.
	\newblock {\em Computer Methods in Applied Mechanics and Engineering},
	198(37-40):2898--2910, 2009.
	
	\bibitem{sun2014arbitrary}
	W Sun, L Wang, T Wu, and G Liu.
	\newblock An arbitrary lagrangian--eulerian model for modelling the
	time-dependent evolution of crevice corrosion.
	\newblock {\em Corrosion Science}, 78:233--243, 2014.
	
	\bibitem{sethian1996fast}
	JA Sethian.
	\newblock A fast marching level set method for monotonically advancing fronts.
	\newblock {\em Proceedings of the National Academy of Sciences},
	93(4):1591--1595, 1996.
	
	\bibitem{mai2016phase}
	W Mai, S Soghrati, and RG Buchheit.
	\newblock A phase field model for simulating the pitting corrosion.
	\newblock {\em Corrosion Science}, 110:157--166, 2016.
	
	\bibitem{yang2021explicit}
	C Yang, H Huang, W Liu, J Wang, J Wang, HM
	Jafri, Y Liu, G Han, H Song, and LQ Chen.
	\newblock Explicit dynamics of diffuse interface in phase-field model.
	\newblock {\em Advanced Theory and Simulations}, 4(1):2000162, 2021.
    
    \bibitem{yang2020multiphase}
    C Yang, X Wang, J Wang, and H Huang.
    \newblock Multiphase-field approach with parabolic approximation scheme.
    \newblock {\em Computational Materials Science}, 172:109322, 2020.
        
    \bibitem{ansari2018phase}
    TQ Ansari, Z Xiao, S Hu, Y Li, JL Luo, and
    SQ Shi.
    \newblock Phase-field model of pitting corrosion kinetics in metallic
    materials.
    \newblock {\em npj Computational Materials}, 4(1):1--9, 2018.	
     
    \bibitem{li2017review}
    Y Li, S Hu, X Sun, and M Stan.
    \newblock A review: applications of the phase field method in predicting
    microstructure and property evolution of irradiated nuclear materials.
    \newblock {\em npj Computational Materials}, 3(1):1--17, 2017.	 
     
    \bibitem{mai2017phase}
    W Mai and S Soghrati.
    \newblock A phase field model for simulating the stress corrosion cracking
    initiated from pits.
    \newblock {\em Corrosion Science}, 125:87--98, 2017.
    
    \bibitem{yang2021hydride}
    C Yang, Y Liu, H Huang, S Wu, J Sheng, X Shi, J Wang,
    G Han, and H Song.
    \newblock Hydride corrosion kinetics on metallic surface: a multiphase-field
    modeling.
    \newblock {\em Materials Research Express}, 8(10):106518, 2021. 
    
    \bibitem{kim1999phase}
    SG Kim, WT Kim, and T Suzuki.
    \newblock Phase-field model for binary alloys.
    \newblock {\em Physical Review E}, 60(6):7186, 1999.
     	
	
	\bibitem{scott2007ud3}
	TB Scott, GC Allen, I Findlay, and J Glascott.
	\newblock UD$_{3}$ formation on uranium: evidence for grain boundary precipitation.
	\newblock {\em Philosophical Magazine}, 87(2):177--187, 2007.
	
	\bibitem{banos2017investigation}
	AK Banos.
	\newblock {\em Investigation of uranium corrosion in mixed water-hydrogen
		environments}.
	\newblock PhD thesis, University of Bristol, 2017.
	
	\bibitem{guo2008elastoplastic}
	XH~Guo, SQ~Shi, QM~Zhang, and XQ~Ma.
	\newblock An elastoplastic phase-field model for the evolution of hydride
	precipitation in zirconium. part \uppercase\expandafter{\romannumeral1}: Smooth specimen.
	\newblock {\em Journal of Nuclear Materials}, 378(1):110--119, 2008.
	
	
	\bibitem{abubakar2015phase}
	AA Abubakar, SS Akhtar, and AFM Arif.
	\newblock Phase field modeling of V$_2$O$_5$ hot corrosion kinetics in thermal
	barrier coatings.
	\newblock {\em Computational Materials Science}, 99:105--116, 2015.
	
	\bibitem{harker2013altering}
	NJ~Harker, TB~Scott, CP~Jones, JR~Petherbridge, and J~Glascott.
	\newblock Altering the hydriding behaviour of uranium metal by induced oxide
	penetration around carbo-nitride inclusions.
	\newblock {\em Solid State Ionics}, 241:46--52, 2013.
	
	\bibitem{stitt2015effects}
	CA~Stitt, C~Paraskevoulakos, NJ~Harker, CP~Jones, and TB~Scott.
	\newblock The effects of metal surface geometry on the formation of uranium
	hydride.
	\newblock {\em Corrosion Science}, 98:63--71, 2015.
	
	\bibitem{steinbach2006multi}
	I Steinbach and M Apel.
	\newblock Multi phase field model for solid state transformation with elastic
	strain.
	\newblock {\em Physica D: Nonlinear Phenomena}, 217(2):153--160, 2006.
	
	\bibitem{powell1973mass}
	GL~Powell and JB~Condon.
	\newblock Mass spectrographic determination of hydrogen thermally evolved from
	uranium and uranium alloys.
	\newblock {\em Analytical Chemistry}, 45(14):2349--2354, 1973.

	\bibitem{bair2017formation}
	J Bair, MA Zaeem, and D Schwen.
	\newblock Formation path of $\delta$ hydrides in zirconium by multiphase field
	modeling.
	\newblock {\em Acta Materialia}, 123:235--244, 2017.
	
	\bibitem{morrell2013uranium}
	JS Morrell, MJ Jackson, et~al.
	\newblock {\em Uranium Processing and Properties}.
	\newblock Springer, 2013.
	
	\bibitem{chiotti1980hu}
	P~Chiotti.
	\newblock The H-U (hydrogen-uranium) system.
	\newblock {\em Bulletin of Alloy Phase Diagrams}, 1(2):99--106, 1980.
	
	\bibitem{morris2010khachaturyan}
	JW~Morris~Jr.
	\newblock The khachaturyan theory of elastic inclusions: Recollections and
	results.
	\newblock {\em Philosophical Magazine}, 90(1-4):3--35, 2010.
	
	\bibitem{yeon2005phase}
	DH Yeon, PR Cha, JH Kim, M Grant, and JK Yoon.
	\newblock A phase field model for phase transformation in an elastically
	stressed binary alloy.
	\newblock {\em Modelling and Simulation in Materials Science and Engineering},
	13(3):299, 2005.
	
	\bibitem{lubliner2008plasticity}
	J~Lubliner.
	\newblock Plasticity theory, revised ed. ed, 2008.
	
	\bibitem{han2019phase}
	GM~Han, YF~Zhao, CB~Zhou, De-Ye Lin, XY~Zhu, J~Zhang, SY~Hu, and HF~Song.
	\newblock Phase-field modeling of stacking structure formation and transition
	of $\delta$-hydride precipitates in zirconium.
	\newblock {\em Acta Materialia}, 165:528--546, 2019.
	
	\bibitem{zhang2005phase}
	JX~Zhang and LQ~Chen.
	\newblock Phase-field microelasticity theory and micromagnetic simulations of
	domain structures in giant magnetostrictive materials.
	\newblock {\em Acta Materialia}, 53(9):2845--2855, 2005.
	
	\bibitem{fisher1966temperature}
	ES~Fisher.
	\newblock Temperature dependence of the elastic moduli in alpha uranium single
	crystals, part iv (298$^{\circ}$ to 923$^{\circ}$ K).
	\newblock {\em Journal of Nuclear Materials}, 18(1):39--54, 1966.
	
	\bibitem{taylor2009ab}
	CD Taylor and RS Lillard.
	\newblock Ab-initio calculations of the hydrogen-uranium system: Surface
	phenomena, absorption, transport and trapping.
	\newblock {\em Acta Materialia}, 57(16):4707--4715, 2009.
	
	\bibitem{bair2016phase}
	J Bair, MA Zaeem, and M Tonks.
	\newblock A phase-field model to study the effects of temperature change on
	shape evolution of $\gamma$-hydrides in zirconium.
	\newblock {\em Journal of Physics D: Applied Physics}, 49(40):405302, 2016.
	
	
	\bibitem{taylor2008evaluation}
	CD Taylor.
	\newblock Evaluation of first-principles techniques for obtaining materials
	parameters of $\alpha$-uranium and the (001) $\alpha$-uranium surface.
	\newblock {\em Physical Review B}, 77(9):094119, 2008.
	
	\bibitem{greenbaum2008strain}
	Y~Greenbaum, D~Barlam, MH~Mintz, and RZ~Shneck.
	\newblock The strain energy and shape evolution of hydrides precipitated at
	free surfaces of metals.
	\newblock {\em Journal of Alloys and Compounds}, 452(2):325--335, 2008.
	
	\bibitem{greenbaum2011elastic}
	Y~Greenbaum, D~Barlam, MH~Mintz, and RZ~Shneck.
	\newblock Elastic fields generated by a semi-spherical hydride particle on a
	free surface of a metal and their effect on its growth.
	\newblock {\em Journal of Alloys and Compounds}, 509(9):4025--4034, 2011.
	
	
	\bibitem{eshelby1957determination}
	JD Eshelby.
	\newblock The determination of the elastic field of an ellipsoidal inclusion,
	and related problems.
	\newblock {\em Proceedings of the Royal Society of London. Series A. Mathematical and Physical Sciences}, 241(1226):376--396, 1957.

	
	\bibitem{evans1969strength}
	AG~Evans and RW~Davidge.
	\newblock The strength and fracture of stoichiometric polycrystalline UO$_{2}$.
	\newblock {\em Journal of Nuclear Materials}, 33(3):249--260, 1969.
	
	\bibitem{harker2012corrosion}
	NJ~Harker.
	\newblock {\em The corrosion of uranium in sealed environments containing
		oxygen and water vapour}.
	\newblock PhD thesis, University of Bristol, 2012.
	

	
\end{thebibliography}
\end{document}


\title{Supplementary Materials to ``A phase-field model for simulating hydrogen-induced pitting corrosion with solid-solid phase transformation in the metal''}

\author{Jie Sheng}
\affiliation{Laboratory of Computational Physics, Institute of Applied Physics and Computational Mathematics, Beijing 100088, China}

\author{Yue-Chao Wang}
\affiliation{Laboratory of Computational Physics, Institute of Applied Physics and Computational Mathematics, Beijing 100088, China}

\author{Yu Liu}
\email{liu\_yu@iapcm.ac.cn}
\affiliation{Laboratory of Computational Physics, Institute of Applied Physics and Computational Mathematics, Beijing 100088, China}

\author{Shuai Wu}
\affiliation{Laboratory of Computational Physics, Institute of Applied Physics and Computational Mathematics, Beijing 100088, China}

\author{Ke Xu}
\affiliation{Laboratory of Computational Physics, Institute of Applied Physics and Computational Mathematics, Beijing 100088, China}

\author{Zi-Hang Chen}
\affiliation{Laboratory of Computational Physics, Institute of Applied Physics and Computational Mathematics, Beijing 100088, China}

\author{Bo Sun}
\affiliation{Laboratory of Computational Physics, Institute of Applied Physics and Computational Mathematics, Beijing 100088, China}

\author{Hai-Feng Liu}
\affiliation{Laboratory of Computational Physics, Institute of Applied Physics and Computational Mathematics, Beijing 100088, China}

\author{Hai-Feng Song}
\email{song\_haifeng@iapcm.ac.cn}
\affiliation{Laboratory of Computational Physics, Institute of Applied Physics and Computational Mathematics, Beijing 100088, China}

\date{\today}
\maketitle

This supplementary materials show the some calculation details of the eigenstrain of the phase transformation among $\alpha$-U, $\alpha$-UH$_{3}$ (metastable phase) and $\beta$-UH$_{3}$, and 
some other simulation results. 
The supplementary materials are organized as follows: Sec. \ref{sec1} shows the crystal structures and corresponding parameters of $\alpha$-U, $\alpha$-UH$_{3}$ and $\beta$-UH$_{3}$. Sec. \ref{sec2} gives the detailed results of orientation change during $\alpha$-U $\rightarrow$ $\alpha$-UH$_{3}$ $\rightarrow$ $\beta$-UH$_{3}$. Sec. \ref{sec3} shows the calculation of the eigenstrain based on the orientation change. Sec. \ref{sec4} shows the simulation about the effect of hydrogen source on pitting morphology.

In our phase-field model, the elastic strain energy is introduced through the eigenstrain, which describes the degree of lattice mismatch and volume expansion during phase transformation\cite{han2019phase}. It can be obtained by analyzing the phase transformation from $\alpha$-U to $\alpha$-UH$_{3}$ and from $\alpha$-UH$_{3}$ to $\beta$-UH$_{3}$. The $\alpha$-UH$_{3}$ is a metastable phase between $\alpha$-U and $\beta$-UH$_{3}$. Some experiments\cite{mulford1954new,banos2018review} has confirmed the existence of $\alpha$-UH$_{3}$ in the phase transformation from $\alpha$-U to $\beta$-UH$_{3}$ at certain temperatures. Therefore, the contribution of $\alpha$-UH$_{3}$ should be considered in the calculation of the eigenstrain.

\section{ Crystal structures of $\alpha$-U, $\alpha$-UH$_{3}$ and $\beta$-UH$_{3}$}\label{sec1}
\begin{figure*}[h]
	\centering
	\includegraphics[width=0.8\textwidth]{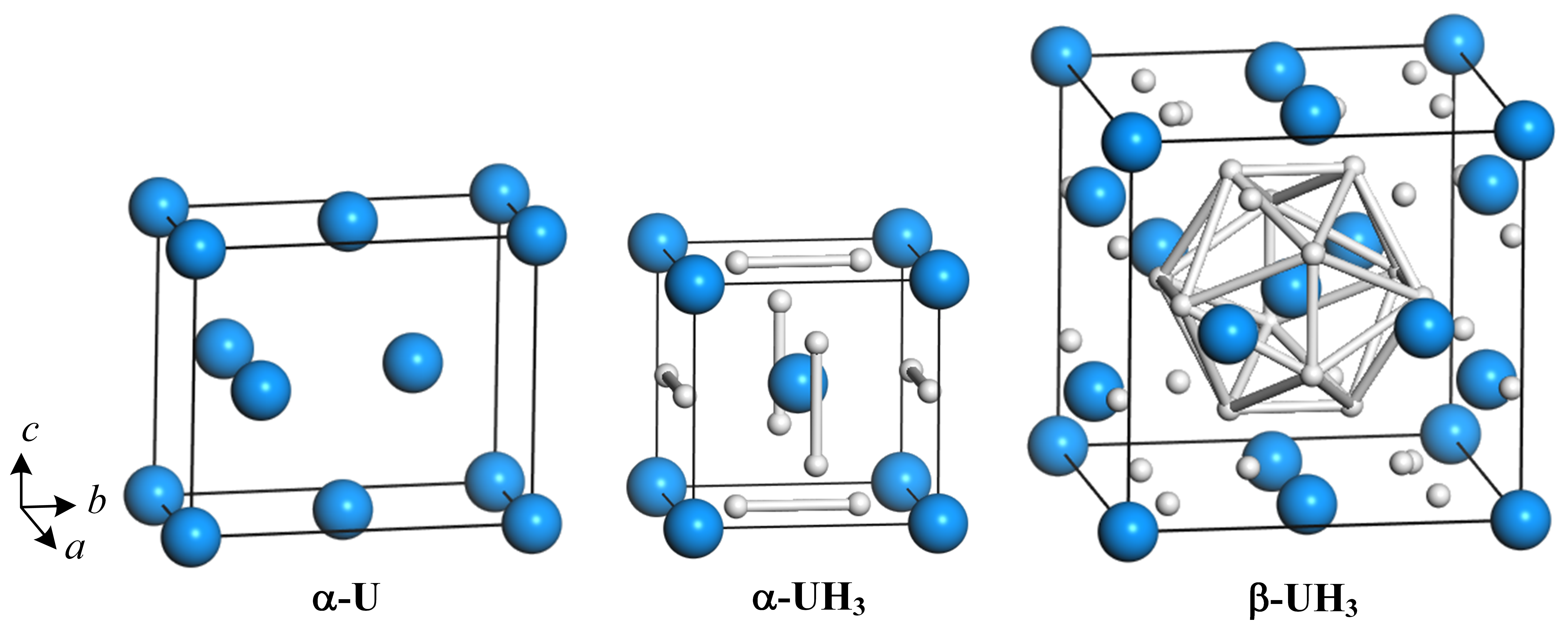}
	\caption{\label{figS1}Crystal structures of $\alpha$-U, $\alpha$-UH$_{3}$ and $\beta$-UH$_{3}$\cite{kruglov2018uranium}. White spheres and blue spheres represent hydrogen and uranium atoms, respectively.}
\end{figure*}

Fig. \ref{figS1} show the  the crystal structures of $\alpha$-U, $\alpha$-UH$_{3}$ and $\beta$-UH$_{3}$\cite{kruglov2018uranium}. Table \ref{tab1} shows the crystal structures parameters of $\alpha$-U, $\alpha$-UH$_{3}$ and $\beta$-UH$_{3}$.

\begin{table}[h]
	\caption{\label{tab1}Crystal structures\cite{banos2018review,lloyd1961thermal} of $\alpha$-U, $\alpha$-UH$_{3}$ and $\beta$-UH$_{3}$.}
	\centering
	\begin{threeparttable}
	\begin{tabular}{|c|c|c|c|c|c|c|}
		\hline
		Phase                            & Space group                   & Lattice parameters                                           & Atom                & x                    & y                    & z                    \\ \hline
		\multirow{3}{*}{$\alpha$-U}      & \multirow{3}{*}{$Cmcm$}       & $a_{\operatorname{\alpha-U}} = 0.288$ nm                     & \multirow{3}{*}{U1} & \multirow{3}{*}{0.0} & \multirow{3}{*}{0.0} & \multirow{3}{*}{0.0} \\
		&                               & $b_{\operatorname{\alpha-U}} = 0.586$ nm                     &                     &                      &                      &                      \\
		&                               & $c_{\operatorname{\alpha-U}} = 0.502$ nm                     &                     &                      &                      &                      \\   \hline
		\multirow{2}{*}{$\alpha$-UH$_{3}$} & \multirow{2}{*}{$Pm\bar{3}n$} & \multirow{2}{*}{$a_{\operatorname{\alpha-UH_{3}}}=0.416$ nm} & U1                  & 0.0                  & 0.0                  & 0.0                  \\
		&                               &                                                              & H1                  & 0.25                 & 0.5                  & 0.0                  \\   \hline
		\multirow{3}{*}{$\beta$-UH$_{3}$}  & \multirow{3}{*}{$Pm\bar{3}n$} & \multirow{3}{*}{$a_{\operatorname{\beta-UH_{3}}} =0.665$ nm} & U1                  & 0.0                  & 0.0                  & 0.0                  \\
		&                               &                                                              & U2                  & 0.25                 & 0.0                  & 0.5                  \\
		&                               &                                                              & H1                  & 0.0                  & 0.1552               & -0.3047             \\
		\hline
	\end{tabular}
\begin{tablenotes}
	\footnotesize
	\item[1] The lattice parameters of $\alpha$-U is measured at temprature 673 K\cite{banos2018review,lloyd1961thermal}.
\end{tablenotes}
\end{threeparttable}
\end{table}

\section{Orientation change during $\alpha$-U $\rightarrow$ $\alpha$-UH$_{3}$ $\rightarrow$ $\beta$-UH$_{3}$}\label{sec2}

\begin{figure*}[h]
	\centering
	\includegraphics[width=0.8\textwidth]{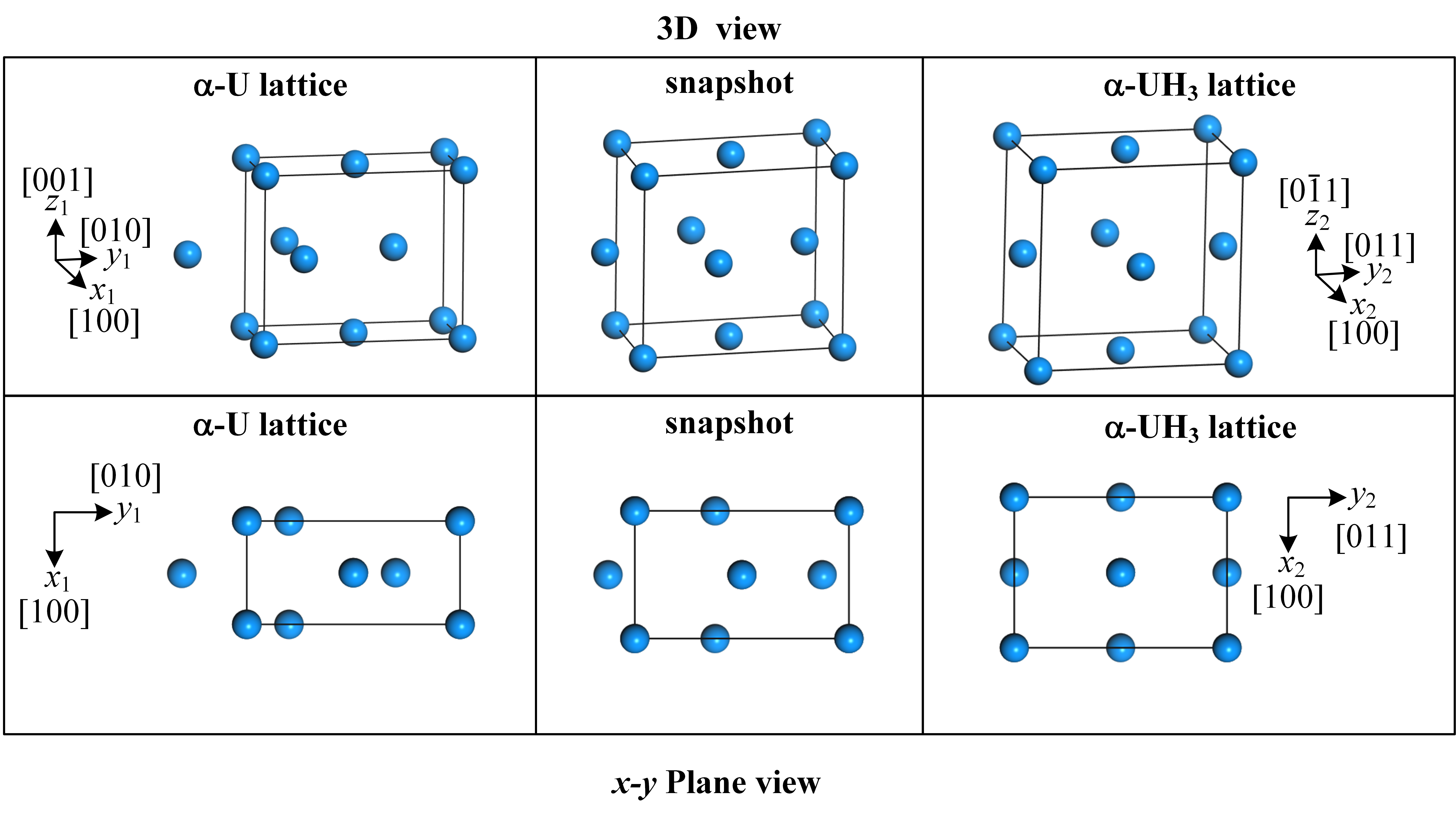}
	\caption{\label{figS2}One of the highest symmetry pathways during the $\alpha$-U to $\alpha$-UH$_{3}$ phase transformation (hydrogen atoms are not shown during the process).}
\end{figure*}

Under different experimental conditions, the movement pathways of atoms may be different, resulting in different orientation changes. According to $\alpha$-U and $\alpha$-UH$_{3}$ structures as described above, Taylor et al.\cite{taylor2010ab} has applied the Bilbao crystallographic research tool to determine symmetry allowed pathways connecting $\alpha$-U with $\alpha$-UH$_{3}$ structures. One of the highest symmetry pathways is shown in Fig. \ref{figS2} (Hydrogen atoms are not shown during the process). The pathway appears to proceed via a ``bunching-up'' of the U atoms in the [0 1 0] direction. The pathway in Fig. \ref{figS2} aligns the [0 1 0] direction of $\alpha$-U with the [0 1 1] direction of $\alpha$-UH$_{3}$.

We have also tried to investigate and analyze the pathway from $\alpha$-UH$_{3}$ to $\beta$-UH$_{3}$. Unfortunately, due to the complexity of $\beta$-UH$_{3}$ structure, it's hard to give a feasible pathway between $\alpha$-UH$_{3}$ and $\beta$-UH$_{3}$ from the first principles calculations. However, based on some physical analysis, we make a general isotropic assumption (that is, the orientation does not change from $\alpha$-UH$_{3}$ to $\beta$-UH$_{3}$). The feasibility of this assumption dependends that the density difference between $\alpha$-UH$_{3}$ (11.12 g/cm$^{3}$) and $\beta$-UH$_{3}$ (10.92 g/cm$^{3}$) is just $1\%$\cite{banos2018review}. Since the variation of material density during the phase transformation affects the elastic strain energy, the contribution of elastic strain energy in this process is far less than that in the process from $\alpha$-U to $\alpha$-UH$_{3}$, which has a density variation about $71\%$  \cite{banos2018review}. In addition, both $\alpha$-UH$_{3}$ and $\beta$-UH$_{3}$ are cubic systems.

\begin{figure*}[h]
	\centering
	\includegraphics[width=0.8\textwidth]{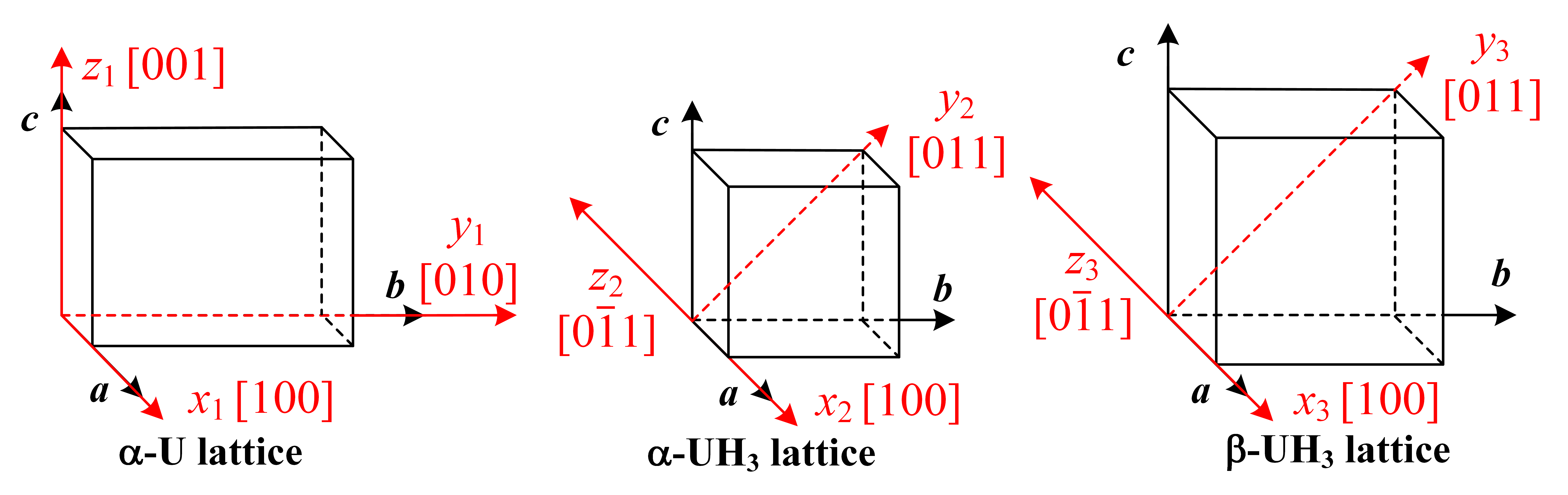}
	\caption{\label{figS3}Orientation and coordinate system selection of $\alpha$-U, $\alpha$-UH$_{3}$ and $\beta$-UH$_{3}$.}
\end{figure*}

According to the pathway and isotropy assumption above, the orientation change of the phase transformation can be given in detail. The orientation and coordinate system selected in this work are shown in Fig. \ref{figS3}. The coordinate system $x_{1}$-$y_{1}$-$z_{1}$ of $\alpha$-U is successively changed into the coordinate system $x_{2}$-$y_{2}$-$z_{2}$ of $\alpha$-UH$_{3}$ and the coordinate system $x_{3}$-$y_{3}$-$z_{3}$ of $\beta$-UH$_{3}$ according to the two steps transformation as:
\begin{align}
	\mathrm{2U^{(\alpha)} + 3H_{2} \rightarrow 2UH_{3}^{(\alpha)}}, \label{eq1}\\
    \mathrm{UH_{3}^{(\alpha)} \rightarrow UH_{3}^{(\beta)}}.\label{eq2}
\end{align}

In the first step (Eq. \eqref{eq1}), based on the coordinate system selected in Fig. \ref{figS3}, the orientation relationship of the phase transformation from $\alpha$-U to $\alpha$-UH$_{3}$ can be described as follows:
\begin{equation}
	\begin{aligned}
		&[1 0 0]_{\operatorname{\alpha-U}} \rm direction \rightarrow [1 0 0]_{\operatorname{\alpha-UH_{3}}} \rm direction: \textit{a}_{\operatorname{\alpha-U}} \rightarrow \textit{a}_{\operatorname{\alpha-UH_{3}}}\nonumber\\
		&[0 1 0]_{\operatorname{\alpha-U}} \rm direction \rightarrow [0 1 1]_{\operatorname{\alpha-UH_{3}}} \rm direction: \textit{b}_{\operatorname{\alpha-U}} \rightarrow \sqrt{2}\textit{a}_{\operatorname{\alpha-UH_{3}}}\nonumber\\
		&[0 0 1]_{\operatorname{\alpha-U}} \rm direction \rightarrow [0 \bar{1} 1]_{\operatorname{\alpha-UH_{3}}} \rm direction: \textit{c}_{\operatorname{\alpha-U}} \rightarrow \sqrt{2}\textit{a}_{\operatorname{\alpha-UH_{3}}}\nonumber			
	\end{aligned}
\end{equation}

 In the second step (Eq. \eqref{eq2}), the coordinate system and orientation from $\alpha$-UH$_{3}$ to $\beta$-UH$_{3}$ do not change under the isotropic assumption, as shown in Fig. \ref{figS3}. In addition, the $\beta$-UH$_{3}$ crystal cell contains four times as many atoms as the $\alpha$-UH$_{3}$ crystal cell, so the volumetric strain of each U atom can be calculated as $\frac{a_{\operatorname{\beta-UH_{3}}}^{3}}{4a_{\operatorname{\alpha-UH_{3}}}^{3}}$. The orientation relation of the phase transformation from $\alpha$-UH$_{3}$ to $\beta$-UH$_{3}$ can be described as follows:
\begin{equation}
	\begin{aligned}
		&[1 0 0]_{\operatorname{\alpha-UH_{3}}} \rm direction \rightarrow [1 0 0]_{\operatorname{\beta-UH_{3}}} \rm direction: \textit{a}_{\operatorname{\alpha-UH_{3}}} \rightarrow \frac{\textit{a}_{\operatorname{\beta-UH_{3}}}}{\sqrt[3]{4}}\nonumber\\
		&[0 1 0]_{\operatorname{\alpha-UH_{3}}} \rm direction \rightarrow [0 1 0]_{\operatorname{\beta-UH_{3}}} \rm direction: \textit{a}_{\operatorname{\alpha-UH_{3}}} \rightarrow \frac{\textit{a}_{\operatorname{\beta-UH_{3}}}}{\sqrt[3]{4}}\nonumber\\
		&[0 0 1]_{\operatorname{\alpha-UH_{3}}} \rm direction \rightarrow [0 0 1]_{\operatorname{\alpha-UH_{3}}} \rm direction: \textit{a}_{\operatorname{\alpha-UH_{3}}} \rightarrow \frac{\textit{a}_{\operatorname{\beta-UH_{3}}}}{\sqrt[3]{4}}			
	\end{aligned}
\end{equation}

\section{Eigenstrain among $\alpha$-U, $\alpha$-UH$_{3}$ and $\beta$-UH$_{3}$}\label{sec3}
Based on the orientation relation described in the previous section, the lattice transformation tensor or deformation gradient tensor $T_1$ and $T_2$, which represent processes from $\alpha$-U to $\alpha$-UH$_{3}$ and from $\alpha$-UH$_{3}$ to $\beta$-UH$_{3}$ respectively, can be calculated by using the finite deformation theory\cite{lubliner2008plasticity}, which can be written as\cite{han2019phase}: 
\begin{align}
	T_1=
	\begin{bmatrix}
		\dfrac{\textit{a}_{\operatorname{\alpha-UH_{3}}}}{\textit{a}_{\operatorname{\alpha-U}}} & 0 & 0\\
		0 &\dfrac{\sqrt{2}\textit{a}_{\operatorname{\alpha-UH_{3}}}}{\textit{b}_{\operatorname{\alpha-U}}} & 0\\
		0 & 0 & \dfrac{\sqrt{2}\textit{a}_{\operatorname{\alpha-UH_{3}}}}{\textit{c}_{\operatorname{\alpha-U}}}\label{eq3}
	\end{bmatrix},
\end{align}
\begin{align}
	T_2=
	\begin{bmatrix}
		\dfrac{\textit{a}_{\operatorname{\beta-UH_{3}}}}{\sqrt[3]{4}\textit{a}_{\operatorname{\alpha-UH_{3}}}} & 0 & 0\\
		0 &\dfrac{\textit{a}_{\operatorname{\beta-UH_{3}}}}{\sqrt[3]{4}\textit{a}_{\operatorname{\alpha-UH_{3}}}} & 0\\
		0 & 0 & \dfrac{\textit{a}_{\operatorname{\beta-UH_{3}}}}{\sqrt[3]{4}\textit{a}_{\operatorname{\alpha-UH_{3}}}}\label{eq4}
	\end{bmatrix}.
\end{align}

Therefore, the total lattice transformation tensor or deformation gradient tensor from $\alpha$-U to $\beta$-UH$_{3}$ is: 
\begin{align}
	U = T_2T_1.\label{eq5}
\end{align}

With Lagrangian finite-strain theory,  the eigenstrain from $\alpha$-U to $\beta$-UH$_{3}$ can be expressed as\cite{lubliner2008plasticity}:
\begin{align}
	\varepsilon_{ij}^{00}=\frac{1}{2}(U^TU-I)= 
	\begin{bmatrix}
		0.5579 & 0       & 0\\
		0      & 0.0111  & 0\\
		0      & 0       &0.1964
	\end{bmatrix}.\label{eq6}
\end{align}
\begin{figure*}[h]
	\centering
	\includegraphics[width=1\textwidth]{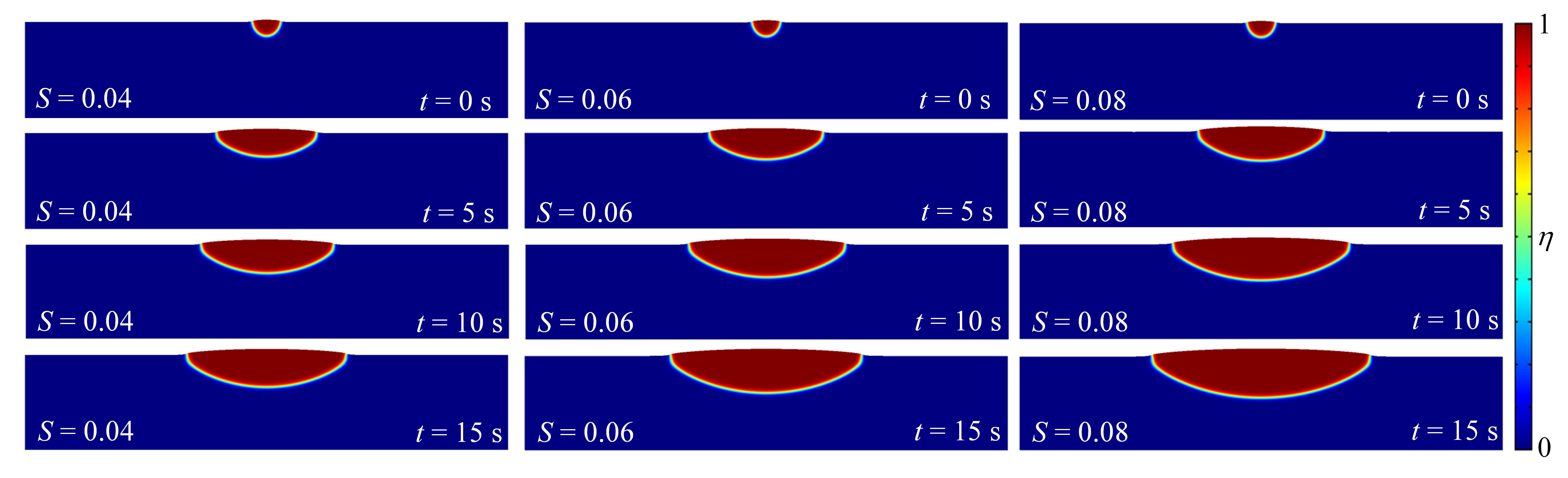}
	\caption{\label{figS4}Pitting morphology under different hydrogen sources ($S$) at different times.}
\end{figure*}
\begin{figure*}[h]
	\centering
	\includegraphics[width=0.5\textwidth]{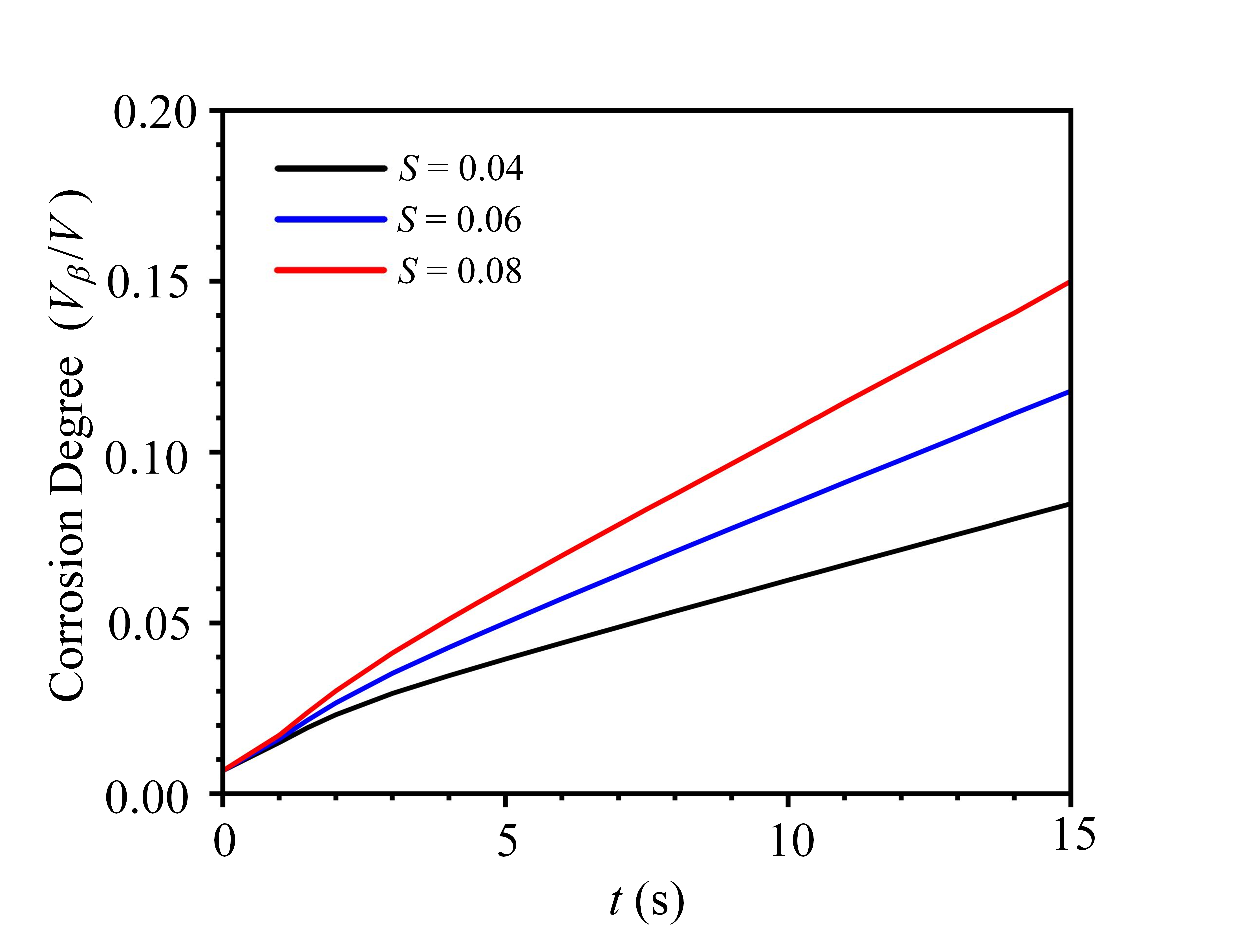}
	\caption{\label{figS5}Corrosion degree under different hydrogen sources ($S$) versus $t$.}
\end{figure*}
where the superscript $T$ represents the matrix transpose and $I$ represents the identity matrix. The results obtained show that the mismatch is anisotropic. This implies a possibility of the ellipsoidal shape of $\beta$-UH$_{3}$ precipitate and anisotropic growth to minimize the deformation energy in the space. 

\section{Effect of hydrogen source on pitting morphology}\label{sec4}
We simulate the pit morphology under different hydrogen sources ($S$) at different times. Fig. \ref{figS4} shows the effect of hydrogen sources on the pit morphology in this PF model. The variation of corrosion degree versus time $t$ under different hydrogen sources is shown in the Fig. \ref{figS5}.

\clearpage